\documentclass[11pt,aps,prc,superscriptaddress,showpacs,floatfix]{revtex4}
\usepackage[dvips]{graphicx}
\usepackage{amsmath,amssymb}
\newcommand{\Nt}{\mbox{${\tilde N}_T$}}
\newcommand{\Nb}{\mbox{${\tilde N}_B$}}
\newcommand{\Nf}{\mbox{${\tilde N}_F$}}
\newcommand{\Et}{\mbox{${\tilde E}_T$}}

\newcommand{\epsi}{\mbox{$\varepsilon$}}

\newcommand{\vx}{\mbox{\boldmath $x$}}

\newcommand{\vbr}{\mbox{\boldmath $r$}}

\newcommand{\vq}{\mbox{\boldmath $q$}}
\newcommand{\vu}{\mbox{\boldmath $u$}}

\newcommand{\vp}{\mbox{\boldmath $p$}}

\newcommand{\vj}{\mbox{\boldmath $j$}}

\newcommand{\Ybfst}{$^{170}$Yb$-^{171}$Yb~}
\newcommand{\Ybsnd}{$^{170}$Yb$-^{173}$Yb~}
\newcommand{\Ybthd}{$^{174}$Yb$-^{173}$Yb~}

\begin{document}

\title{Quadrupole Oscillations in Bose-Fermi Mixtures of Ultracold Atomic Gases 
made of Yb atoms  
\\in the Time-Dependent Gross-Pitaevskii and Vlasov equations}

\author{Tomoyuki~Maruyama}
\affiliation{College of Bioresource Sciences,
Nihon University,
Fujisawa 252-8510, Japan}
\affiliation{Advanced Science Research Center,
Japan Atomic Energy Research Institute, Tokai 319-1195, Japan}

\author{Hiroyuki Yabu}
\affiliation{
Department of Physics,
Ritsumeikan University, Kusatsu 525-8577, Japan}

\date{\today}

\begin{abstract}
We study quadrupole collective oscillations in the bose-fermi mixtures 
of ultracold atomic gases of Yb isotopes, 
which are realized by Kyoto group.
Three kinds of combinations are chosen,
\Ybfst, \Ybsnd and \Ybthd, 
where boson-fermion interactions are weakly repulsive, 
strongly attractive and strongly repulsive respectively.
Collective oscillations in these mixtures are calculated 
in a dynamical time-evolution approach formulated with
the time-dependent Gross-Pitaevskii and the Vlasov equations.
The boson oscillations are shown to have one collective mode, 
and the fermions are shown to have the boson-forced and 
two intrinsic modes, which correspond to the inside- and outside-fermion 
oscillations for the boson-distributed regions.
In the case of the weak boson-fermion interactions, 
the dynamical calculations are shown to be consistent with 
the results obtained in the small amplitude approximations 
as the random phase approximation in early stage of oscillation, 
but, in later stage, these two approaches are shown to give 
the different results.
Also, in the case of the strong boson-fermion interactions, 
discrepancies appear in early stage of oscillation. 
We also analyze these differences in two approaches, 
and show that they originated in the change of the fermion distributions
through oscillation.
\end{abstract}



\maketitle

\section{Introduction}
\label{intr}

Over the last several years, 
there have been significant progresses 
in ultracold atomic gas physics \cite{GPS}: 
Bose-Einstein condensates (BEC) \cite{nobel,Dalfovo,becth,Andersen}, 
two boson mixtures  \cite{B-B}, 
Fermi-degenerate atomic gases \cite{ferG},
and Bose-Fermi (BF) mixtures \cite{Schreck,BferM,Modugno}, 
and so on.   
In particular,
the BF mixtures attract physical interests as a typical example 
where particles obeying different quantum statistics are intermingled.
In the study of this system, 
we have a big opportunity to obtain
a lot of new knowledge on quantum many-body systems 
because we can make a variety of mixtures with atomic specie combinations 
and can control the atomic interactions 
using the Feshbach-resonance method \cite{Fesh}.
Theoretical studies of the BF mixtures have been done 
on static properties \cite{Molmer,Amoruso,MOSY,Bijlsma,Vichi,Vichi1}, 
the phase structures and separation \cite{Nygaard,Yi,Viverit,Capuzzi02}, 
induced instabilities by the attractive interactions \cite{MSY1,Roth,Capuzzi03},
and the collective excitations
\cite{MSY,Minguzzi,zeros,Yip,sogo,miyakawa,Liu,tomoBF,monoEX,tbDPL,scalTF,scalTF2}. 

One of important diagnostic signals 
of many-particle systems
is the collective excitations
because of their sensitivity 
on the inter-atomic interactions 
and the ground- and excited-state structures.  
Theoretically,
collective motions are usually studied 
in the random phase approximation (RPA) \cite{zeros,sogo} 
or its approximate methods: 
the sum-rule \cite{miyakawa,MSY}
or the scaling \cite{Liu,scal,scalDP,scalTF,scalTF2} methods.

In the previous papers, 
we calculated time-evolution of the BF mixtures directly
with solving the time-dependent Gross-Pitaevskii (TDGP) and the Vlasov equations, 
and studied their monopole \cite{tomoBF,monoEX} and dipole \cite{tbDPL} oscillations;
the results are largely consistent with the RPA calculations \cite{sogo}, 
but show different behaviors in some aspects:
for example, the rapid damping at zero temperature ($T=0$).

In the BF mixtures at  $T=0$, 
the condensed bosons occupy one single-particle state,  
and the fermions distribute in a wide range of single-particle states.
Thus the boson oscillation has only one collective mode in spectrum
with no strong damping \cite{chevy,tohyama};
on the other hand,
various collective modes appear  
in the fermion collective oscillation generally.
The RPA calculations \cite{ColBTh} can explain the experimental results 
on the frequencies of collective motions\cite{ColEx}.
Especially, in the fermion oscillation of the BF mixtures,  
two regions of the fermi gas, inside and outside of the boson distributions,
oscillate with different frequencies, 
and their interference gives the beat and damping phenomena, 
which are clearly shown 
in the time-evolution approach; 
on the other hand, the boson oscillation become monotonous \cite{tomoBF,tbDPL}. 

In the time-evolution approach,
the intrinsic frequencies of the oscillation modes can be obtained 
from the Fourier transform of the time-dependence 
of the collective coordinates.
In the dipole oscillation of the fermion components,
we have really confirmed the existence 
of the two intrinsic modes corresponding to the inside- and outside motions
and one forced-oscillation mode caused by the boson oscillation \cite{tbDPL}.
In the early stage of the oscillation, 
these intrinsic frequencies obtained in the time-evolution approach are consistent 
with the RPA calculation, 
but, in the later stage, the two calculations show different results \cite{tomoBF}; 
the difference is originated in the distribution changes of the density and the velocity 
in time-developments, 
which appears in the time-evolution approach 
but not in harmonic approximations like the RPA calculations. 
In actual experiments, 
the oscillation amplitude is not so small \cite{ColEx} that 
the results obtained in the time-dependent approach 
should be more reliable 
in comparison with experiments.

In this paper, we consider the BF mixtures of the Yb isotopes, 
which have some particular properties; 
the Yb consists of many kinds of isotopes, 
five bosons (${}^{168,170,172,174,176}$Yb) 
and two fermions (${}^{171,173}$Yb), 
which give a variety of combinations 
in the BF mixtures.
Experimental researches on the trapped atomic gases of the Yb isotopes 
are being performed actively by the group of Kyoto university; 
the BEC \cite{Takasu} and the Fermi-degeneracy \cite{Fukuhara1} have been performed.  
The scattering lengths for the boson-fermion interactions have been obtained 
experimentally by the group \cite{Fukuhara1,Fukuhara2,Kitagawa}, 
and the observation of the ground state properties and the collective oscillations 
of the BF mixtures is now under progressing. 

In this paper, 
we discuss the quadrupole oscillations of the BF mixtures of Yb isotopes 
in the time-evolution approach, 
where the time evolution of the condensed-boson wave function
and the fermion phase-space distribution function 
are obtained from the solutions 
of the TDGP and Vlasov equations, respectively. 
In the next section, we give the formulation of the transport model
to calculate the time evolution. 
In Sec. III, 
the numerical results on the quadrupole oscillations  
are shown with their physical properties 
for three kinds  of BF mixtures, \Ybfst, \Ybsnd and \Ybthd.
Sec. IV is for summary.

%
%
\section{Time-Dependent Gross-Pitaevskii and Vlasov Equations}
\label{TevEq}
%
%

In this section, we briefly explain the time-evolution approach 
to calculate collective oscillations of the BF mixture.
Let's consider the system of the coexistent dilute gases 
of one bosonic and one-component fermionic atoms at $T=0$, 
which is trapped in the axially-symmetric potential with respect to the $z$-axis. 
The zero-range boson-boson and boson-fermion interactions are assumed, 
and no fermion-fermion interaction exists in the system. 
Then the hamiltonian of the system is 
\begin{eqnarray}
     {\tilde H} = \int d^3{q}~\Bigg[
         &-& \frac{\hbar^2}{2 M_B} {\tilde \phi}^\dagger(\vq) \nabla^2_q {\tilde \phi}(\vq)
          +  \frac{1}{2} M_B \Omega_B^2 (\vq_T^2 + \kappa_L^2 q_L^2) 
             {\tilde \phi}^\dagger(\vq) {\tilde \phi}(\vq) 
\nonumber \\
         &+& \frac{2 \pi \hbar^2 a_{BB}}{M_B} 
             \{ {\tilde \phi}^\dagger (\vq) {\tilde \phi} (\vq) \}^2
\nonumber \\
         &-& \frac{\hbar^2}{2M_f} {\tilde \psi}^\dagger (\vq) \nabla_q^2 {\tilde \psi} (\vq)
          + \frac{1}{2} M_f  \Omega_F^2  (\vq_T^2 + \kappa_L^2 q_L^2) 
            {\tilde \psi}^\dagger(\vq) {\tilde \psi} (\vq)
\nonumber \\
         &+& \frac{2 \pi \hbar^2 a_{BF}(M_B + M_F)}{M_B M_F}
             {\tilde \phi}^\dagger (\vq) {\tilde \phi} (\vq) 
             {\tilde \psi}^\dagger (\vq) {\tilde \psi} (\vq) \Bigg],
\label{eq1}
\end{eqnarray}
where ${\tilde \phi}$ and ${\tilde \psi}$ are boson and fermion fields, respectively, 
$M_{B,F}$ are the boson and fermion masses,
$\Omega_{B,F}$ are the transverse frequencies of the trapping potentials 
for the boson and the fermion, 
and $a_{BB,BF}$ are the boson-boson and boson-fermion $s$-wave scattering lengths.
The transverse and longitudinal components 
of the spatial coordinate are described by $\vq \equiv (\vq_T, q_L)$,
and $\kappa_L$ is the longitudinal-to-transverse frequency ratio
of the trapping potentials; 
in this paper, 
we assume the same ratio for the bosons and fermions.

To reduce the parameters, 
we rewrite Eq.~(\ref{eq1}) with the dimensionless variables;
the scaled spatial coordinates $R_B = (\hbar / M_B \Omega_B)^{1/2}$ 
and the scaled boson/fermion fields 
$\phi = R_B^{-1/3} {\tilde \phi}$ \& $\psi = R_B^{-1/3} {\tilde \psi}$, 
where the scaling parameter $R_B$ is defined by 
$R_B = (\hbar / M_B \Omega_B)^{1/2}$.
Then, the scaled hamiltonian $H \equiv {\tilde H} / \hbar \Omega_B$ becomes
\begin{eqnarray}
     H = \int d^3{r}~\Bigg[
          &-& \frac{1}{2} \phi^\dagger(\vbr) \nabla^2_r \phi(\vbr)
           +  \frac{1}{2} (\vbr_T^2 + \kappa_L^2 z^2) \phi^\dagger(\vbr) \phi(\vbr) 
\nonumber \\
          &+& \frac{g_{BB}}{2} \{ \phi^\dagger(\vbr) \phi (\vbr) \}^2 
\nonumber \\
          &-& \frac{1}{2m_f} \psi^\dagger(\vbr) \nabla^2_r \psi (\vbr)
           +  \frac{1}{2} m_f \omega_f^2 (\vbr_T^2 + \kappa_L^2 z^2) 
                         \psi^\dagger(\vbr) \psi (\vbr) 
\nonumber \\
          &+& h_{BF} \phi^\dagger(\vbr) \phi (\vbr) 
                     \psi^\dagger(\vbr) \psi (\vbr) \Bigg],
\label{TotH}
\end{eqnarray}
where the dimensionless parameters are defined as 
$m_f \equiv M_F/M_B$ (fermion mass),  
$\omega_f \equiv \Omega_F/\Omega_B$ (fermion trapping-potential frequency), 
$g_{BB} \equiv 8 \pi \hbar a_{BB} R_B^{-1}$ and 
$h_{BF} \equiv 4 \pi \hbar m_f a_{BF} (1+m_f)^{-1} R_B^{-1}$ 
(boson-boson and boson-fermion coupling constants).

In this paper, we consider the $T=0$ system
with $N_b$ bosons and $N_f$ fermions;
the total wave function of the system is written by
\begin{equation}
     \Phi(\tau) = \left\{\prod_{i=1}^{N_b} \phi_c (\vbr_i) \right\} \Psi_f[\psi_n], 
\end{equation}
where $\phi_c$ is the condensed-boson wave function, and
$\Psi_f$ is the fermion many-body wave function, 
which is given by the Slater determinant 
of fermion single-particle wave functions $\psi_n$,
where $n$ is the quantum number.
  
The time-evolution equations of the total wave function $\Phi(\tau)$ is obtained 
from the variational condition: 
\begin{equation}
     \delta \int{d\tau} 
          \langle \Phi(\tau) | i\frac{\partial}{\partial \tau} - H |\Phi(\tau) \rangle = 0.
\label{vari}
\end{equation}
It gives the coupled TDGP and TDHF equations:
\begin{eqnarray}
     i \frac{\partial}{\partial \tau} \phi_c(\vbr,\tau) 
          &=& \left\{ -\frac{1}{2} \nabla_r^2 +U_B(\vbr) \right\} \,\phi_c(\vbr,\tau),
\label{TDGP}\\
     i \frac{\partial}{\partial \tau} \psi_n(\vbr,\tau) 
          &=& \left\{ -\frac{1}{2 m_f} \nabla_r^2 +U_F(\vbr) \right\} \,\psi_n(\vbr,\tau).
\label{TDHF}
\end{eqnarray}
The effective potentials $U_B$ and $U_F$ are
\begin{eqnarray}
     U_B(\vbr) &=& \frac{1}{2} (\vbr_T^2 +\kappa_L^2 z^2)
                +g_{BB} \rho_B(\vbr) + h_{BF} \rho_F(\vbr),
\label{uB}
\\
     U_F(\vbr) &=& \frac{1}{2} m_f \omega^2_f (\vbr_T^2 + \kappa_L^2 z^2)
                +h_{BF} \rho_B(\vbr) ,
\label{uF}
\end{eqnarray}
where $\rho_B$ and $\rho_F$ are the boson and fermion densities:
\begin{eqnarray}
     \rho_B(\vbr) &=& N_b |\phi_c(\vbr)|^2,
\label{rhoB}\\
     \rho_F(\vbr) &=& \sum^{occ}_{n} |\psi_n (\vbr)|^2.
\label{rhoF}
\end{eqnarray} 

The number of fermion occupied states in the above equations are usually too large 
to solve the above TDHF equations numerically,
so we use the semi-classical approach.
In the semi-classical limit ($\hbar \rightarrow 0$), 
the TDHF equation is proved to be equivalent with
the Vlasov equation \cite{KB}:
\begin{equation}
     \frac{d}{d \tau} f(\vbr,\vp;\tau) =
          \left\{ \frac{\partial}{\partial \tau} 
                 +\frac{\vp}{m_f}{\nabla_r}
                 -[\nabla_r U_F(\vbr)][\nabla_p] \right\} f(\vbr,\vp;\tau) = 0,
\label{Vlasov}
\end{equation}
where $f(\vbr,\vp;\tau)$ is the fermion phase-space distribution function:
\begin{equation}
     f(\vbr,\vp,\tau) = \int {d^3 u} 
                        \langle \Phi| \psi(\vbr+\frac{1}{2}{\vu},\tau)
                                      \psi^{\dagger}(\vbr-\frac{1}{2}{\vu},\tau) 
                                |\Phi \rangle e^{-i \vp \vu }.
\end{equation}

In actual numerical calculation, 
we use the test particle method \cite{TP}
to solve the Vlasov equation (\ref{Vlasov}).
In this method, the fermion phase-space distribution function is described by
\begin{equation}
     f(\vbr,\vp,\tau) = \frac{(2 \pi)^3}{\Nt} 
          \sum_{i=1}^{{\tilde N}_T N_f} \delta\{\vbr-\vbr_i(\tau)\}
                                        \delta\{\vp-\vp_i(\tau)\},
\label{TP-Wig}
\end{equation}
where {\Nt} is the number of test-particles per fermion.
Substituting Eq.~(\ref{TP-Wig}) into Eq.~(\ref{Vlasov}),
we obtain the equations of motion for the test-particles: 
\begin{eqnarray}
     \frac{d}{d \tau} \vbr_i (\tau) &=& \frac{\vp_i}{m_f},
\label{eqM1} \\
     \frac{d}{d \tau} \vp_i (\tau) &=& - \nabla_r U_F(\vbr_i).
\label{eqM2}
\end{eqnarray}

Thus the time evolutions
of the condensed-boson wave function $\phi_c$ and the fermion phase space
distribution function $f(\vbr,\vp,\tau)$ are obtained 
by solving Eqs.~(\ref{TDGP}), (\ref{eqM1}) and (\ref{eqM2}).

%
%
\section{Ground States and Transition Strengths}
\label{ResGRD}
%
%

Before we show the results on collective excitations in the TDGP+Vlasov approach,
we discuss the ground and excited states of the BF mixtures of Yb isotopes 
in RPA calculation.  

The $s$-wave scattering lengths for the interatomic interactions 
between Yb isotopes have been obtained by Kyoto group \cite{Kitagawa};
these results permit three kinds of BF mixtures,
\Ybfst, \Ybsnd and \Ybthd,  
which are available in the present calculation.
They are also interesting combinations 
because the boson-fermion interactions are 
weakly repulsive (\Ybfst), strongly attractive (\Ybsnd) and strongly repulsive (\Ybthd), 
respectively.
The values of the coupling constants used in the present calculation 
are shown in TABLE I.

For the boson and fermion numbers in the mixtures, 
we take $N_b = 10000$ and $N_f = 1000$,
which are consistent with the experiments by Kyoto group.
In this paper, 
we assume the spherical trapping potentials ($\kappa_L = 1$) 
with the trapping potential frequency $\Omega_B = 2 \pi \times 100\,{\rm Hz}$ 
for bosons and fermions $\omega_f = \Omega_F/\Omega_B =1$.
We also use the approximation $m_f =M_F/M_B=1$ for all kinds of mixtures. 

\subsection{Ground States}
\label{GRD}

The ground-state wave function of the condensed boson $\phi_c^{(g)}$ is obtained 
from the Gross-Pitaevskii equation:
\begin{eqnarray}
     \left\{ - \frac{1}{2} \nabla_r^2 
             + \frac{1}{2} \vbr^2 
             + g_{BB} \rho_B (\vbr) + h_{BF} \rho_F (\vbr) \right\}\,
     \phi^{(g)}_c (\vbr)  = \mu_b \phi_c^{(g)} (\vbr),
\label{GPgr}
\end{eqnarray}
where $\mu_b$ is the boson chemical potential.
To evaluate the fermion phase-space distribution function 
in the ground state, 
we use the Thomas-Fermi (TF) approximation, 
where the distribution function is assumed to have the form: 
\begin{equation}   
     f(\vbr,\vp ) = \theta[ \mu_f - \epsi(\vbr,\vp) ].
\label{TFapp}
\end{equation}
with the fermion chemical potential $\mu_f$.
The phase-space one-particle energy $\epsi(\vbr,\vp)$ in (\ref{TFapp}) is defined by
\begin{equation}
     \epsi(\vbr, \vp) = \frac{1}{2m_f} \vp^2 + U_F (\vbr) ,
\end{equation}
where the effective potential $U_F$ is given by (\ref{uF}).
The fermion density $\rho_F$ included in $U_F$ is 
related with the boson density $\rho_B$ through the Thomas-Fermi equation:
\begin{equation}
     \frac{1}{2 m_f} \left\{6 \pi^2 \rho_F (\vbr) \right\}^{2/3}
          + \frac{1}{2} m_f \omega^2_f (\vbr_T^2 + \kappa_L^2 z^2)
          + h_{BF} \rho_B (\vbr) = \mu_f.
\label{TFf}
\end{equation}
Iterating of solving the boson wave function with Eq.~(\ref{GPgr}) 
and searching the fermi energy $\mu_f$ in Eq.~(\ref{TFf}) 
for the correct fermion number, 
we obtain the ground-state wave function $\phi^{(g)}$ and the fermion density distribution $\rho_F$ 
to the ground states of the BF mixtures.

\begin{figure}[ht]
\hspace*{0cm}
{\includegraphics[scale=0.65]{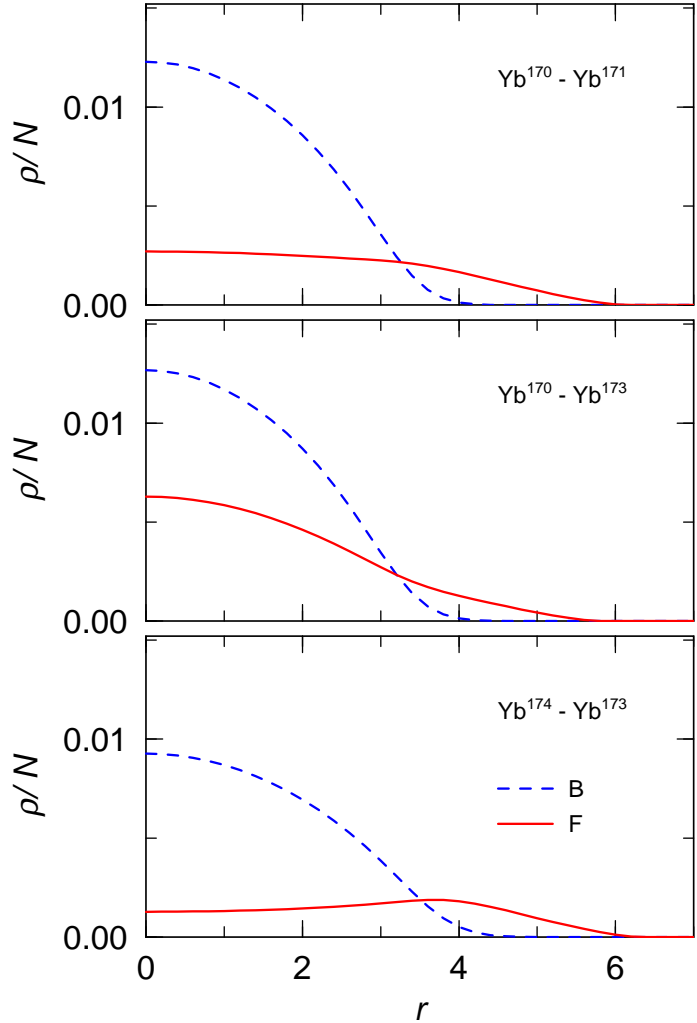}}
\caption{\small
The ground-state density distributions of the BF mixtures:
(a) $^{170}$Yb$-^{171}$Yb, (b) $^{170}$Yb$-^{173}$Yb 
and (c) $^{174}$Yb$-^{173}$Yb (c).
The dashed and solid lines represent the density of
the bosons and fermions, respectively.}
\label{grdRH}
\end{figure}

Thus obtained boson and fermion density distributions are shown in Figs.~\ref{grdRH}a-c. 
In the mixtures of  {\Ybfst} (Fig.\ref{grdRH}a) and {\Ybsnd} (Fig.\ref{grdRH}b),
the density distributions are center-peaked,
but, in the \Ybsnd mixture, 
the fermion distribution has large overlap
with the boson density, 
because of the attractive boson-fermion interaction ($h_{BF} < 0$).
On the other hand, 
in the \Ybthd mixture (Fig.\ref{grdRH}c),
the fermion density distribution is surface-peaked;
it is because the BF interaction is strong enough 
to satisfy the condition $h_{BF}/g_{BB} > 1$ \cite{miyakawa}.
The $h_{BF}$-dependence of the fermion density distribution 
in the TF approximation 
is explained in Appendix \ref{app1}, 
where the ground-state density profiles are shown to be determined 
by the single parameter $h_{BF}/g_{BB}$. 

\subsection{Excitation Energies and Transition Strengths in RPA}

As shown in the preceding calculations \cite{miyakawa,tbDPL}, 
the collective-oscillation frequency is sensitive to the profiles 
of the density distributions.
Here we calculate the excited states in the random phase approximation (RPA), 
which is a good approximation in the case of the small amplitude, 
and show the transition strengths between the excited and ground states.

For this purpose,
we calculate the fermion ground-state wave functions 
in the Hartree-Fock (HF) approximation and, then, 
the excited states are obtained using the ground state in RPA 
as in Ref.~\cite{sogo}.

In the HF approximation, 
the fermion number $N_f$ is restricted to the values 
that are determined from the subshell closure of fermion single-particle states, 
so that it become slightly different from the value of the present calculation;
however, it does not affect the final results.
The boson and fermion transition amplitudes from an excited state
$|\Phi_n \rangle$ with the excitation energy $\omega_n$ 
to the ground state $|\Phi_0 \rangle$ 
are defined by 
\begin{eqnarray}
     T_B(\omega_n) &=& \langle \Phi_n| {\hat O}_B |\Phi_0 \rangle,
\label{ABn}\\
     T_F(\omega_n) &=& \langle \Phi_n| {\hat O}_F |\Phi_0\rangle,
\label{AFn}
\end{eqnarray}
where the quadrupole operators ${\hat O}_{B,F}$ are 
\begin{eqnarray}
     {\hat O}_B &=& \int d^3{r} \phi^*(\vbr) (\vbr_T^2 -2 z^2) \phi(\vbr), 
\label{qpOB}\\
     {\hat O}_F &=&  \int d^3{r} \psi^*(\vbr) (\vbr_T^2 -2 z^2) \psi(\vbr).
\label{qpOF}\\
\end{eqnarray}

\begin{figure}[ht]
\hspace*{0cm}
{\includegraphics[scale=0.68,angle=270]{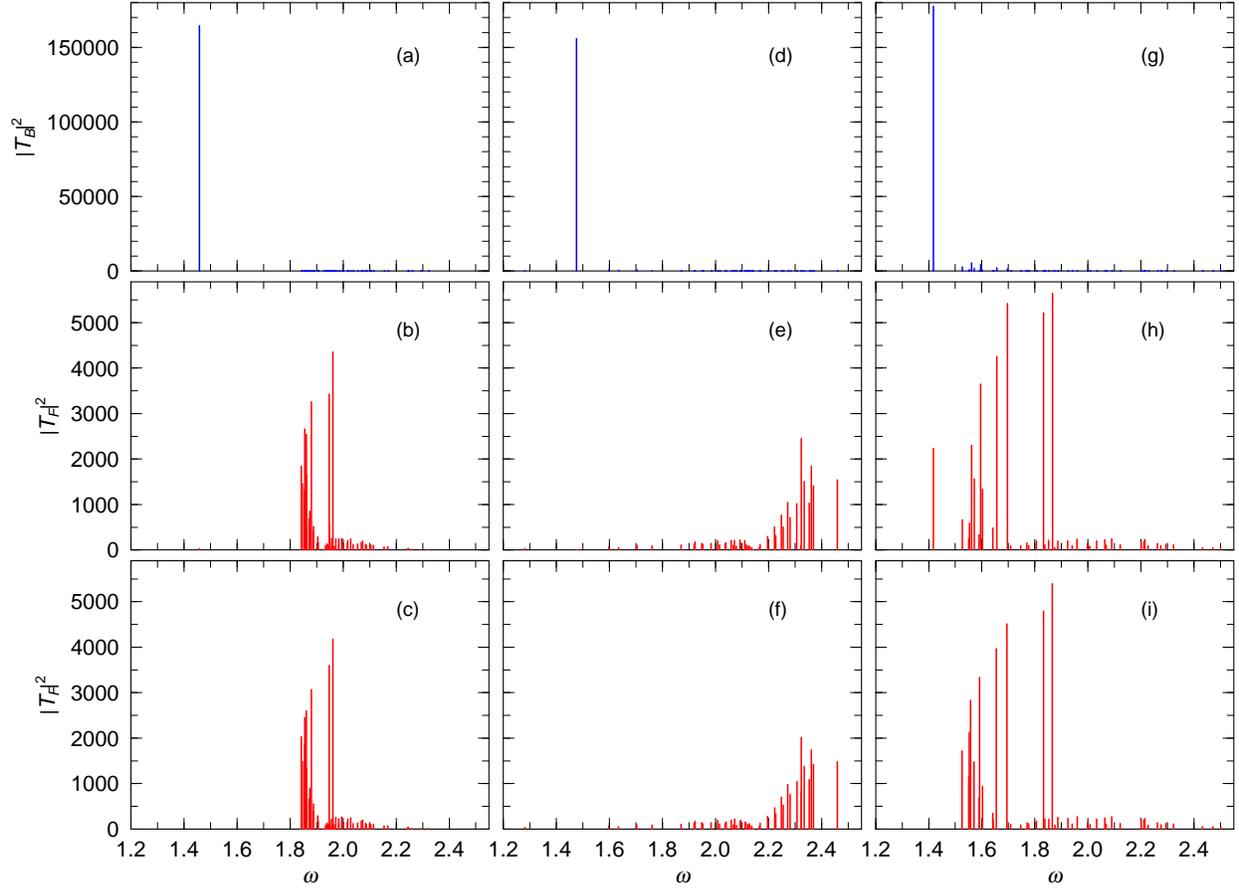}}
\caption{\small
The transition strengths $|A_{B,F}|^2$ in RPA: 
\Ybfst (left panels), \Ybsnd (center panels) and \Ybthd (right panels). 
The top, middle and bottom panels are for the boson, fermion and 
the single-particle fermion strengths}
\label{TstRPA}
\end{figure}

In Fig.~\ref{TstRPA}, 
the transition strengths $|T_{B,F}|^2$ in RPA are shown 
for   
\Ybfst (left panels), \Ybsnd (center panels) and \Ybthd (right panels)
mixtures, 
where the boson and fermion transition strengths are plotted 
in the top and middle panels, respectively.
In the bottom panels, 
the fermion strengths in the single particle processes are added.

As is clearly shown in the top panels, 
the boson transition strength $|T_B(\omega)|^2$ is almost concentrated on one excited state, 
whose excitation energy is denoted by $\omega_Q^b$. 
On the other hand, 
the fermion transition strength $|T_F(\omega)|^2$ have two peaks,
and distribute between them (middle panels).
In \Ybfst (left panels) and \Ybsnd (center panels),
the fermion strength $|T_F(\omega)|^2$ shows no sharp peaks 
at the boson excitation energy of $\omega_Q^b$;
it has almost the same structure as the strength
in the single particle process (bottom panels).
In contrast,
the \Ybthd mixture shows a visible strength at $\omega_Q^b$ (Fig.~\ref{TstRPA}h), 
which is very small in the single particle process (Fig.~\ref{TstRPA}i).

The two peaks of the fermion transition strength 
in the \Ybfst correspond to the fermion intrinsic modes. 
As shown in Ref. \cite{tbDPL},
the fermion transition strength of the dipole oscillation 
also shows two peaks, 
which correspond to the fermion motions in the inside/outside regions of the bosons
(the inside and outside fermion oscillations).    
Indeed, 
the second-peak frequency of the quadrupole oscillation 
is $\omega \approx 2$:
the value in the pure harmonic oscillator potential.
According to Ref.~\cite{tbDPL}, 
we call these two modes the mode-3 (the inside-fermion oscillation) 
and mode-2 (the outside-fermion oscillation), respectively;
then, we define the fermion intrinsic frequency $\omega_Q^f$ 
as the peak position of the mode-3 in the fermion strength function $|T_F(\omega)|^2$.
As shown in Fig.~\ref{TstRPA}h, 
the $|T_F(\omega)|^2$ includes an additional mode.
From  Fig.~\ref{TstRPA}g, 
we can find that it correspond to the oscillation forced 
by the boson intrinsic oscillation;
we call it ``the boson-forced oscillation'' (mode-1).

It should be noted that, 
in the \Ybsnd and \Ybthd mixtures, 
the fermion strength functions have two peaks
(except the peak at $\omega=\omega_Q^b$)
but no clear peaks at $\omega \approx 2$ corresponding to the mode-2 (Figs.~\ref{TstRPA}e,h).
It is because one of the two peaks (the stronger peak) are broad in these mixtures, 
and the two peaks are merged into one broad peak. 

%
%
\section{Quadrupole Oscillations of Boson-Fermion Mixtures of Yb Isotopes}
\label{ResD}
%
%

In this section, we show the results 
of the time-evolution calculations  
for the quadrupole oscillations in the BF mixture.

\subsection{Time-Evolution of Boson-Fermion Mixtures and Description of Quadrupole Oscillations}

As the initial condition of time-evolution at $\tau= 0$, 
we take the boosted state from the ground-state condensed-boson wave function $\phi^{(g)}$ 
and fermion test particle momentum $(\vp_T^{(g)},p_z^{(g)})$, 
which are defined by 
\begin{equation}
     \phi_c(\vbr,\tau=0) = \exp\left\{ \frac{i \lambda_B}{2\sqrt{2}} (r_T^2 - 2 z^2)  \right\} 
                           \phi_c^{(g)} (\vbr) ,
\noindent
\label{bin}
\end{equation}
and 
\begin{equation}
     \vp_T(i) = \vp_T^{(g)}(i) + m_f \omega_f \lambda_F\, \vbr_T,  \qquad
     p_z (i) = p_z^{(g)}(i) - 2 m_f \omega_f \lambda_F z.
\label{fin}
\end{equation}
In Eqs.~(\ref{bin}) and (\ref{fin}), 
$\lambda_B$ and $\lambda_F$ are the boost parameters.
Then, the boson and fermion current densities for the initial state
become
\begin{eqnarray}
     \vj_B(\vbr,\tau=0) &=& \frac{\lambda_B }{\sqrt{2}}
                            (\vbr_T - 2 z {\hat z}) \rho_B^{(g)}(\vbr), \\
     \vj_F(\vbr,\tau=0) &=& \omega_f \lambda_F (\vbr_T - 2 z {\hat z}) \rho_F^{(g)}(\vbr).
\end{eqnarray}

In order to discuss the time-dependence of the quadrupole oscillations,
we define the quantities $Q_{B,F}$ for boson and fermion:
\begin{equation}
     Q_{B,F} = \frac{ R_T^2(B,F) - 2 R_L^2(B,F)}{R_0^2(B,F)},
\label{QBF}
\end{equation}
where
\begin{eqnarray}
     R_T^2(B,F) &=& \frac{1}{N_{b,f}} \int d^3{r}\,\vbr_T^2 \rho_{B,F}(\vbr), 
\label{RT}\\
     R_L^2(B,F) &=& \frac{1}{N_{b,f}} \int d^3{r}\,z^2 \rho_{B,F}(\vbr),
\label{RL}
\end{eqnarray}
and $R_0(B)$ and $R_0 (F)$ are the root-mean-square radii of the boson and
fermion distributions in the ground state.

In the case of no BF interactions ($h_{BF} = 0$), 
the quantities $Q_B$ and $Q_F$ are proportional 
to $\lambda_B \sin(\sqrt{2} \tau)$ and $\lambda_F \sin(2 \omega_f \tau)$, respectively;
they oscillate monotonously with the periods $\sqrt{2}$ and $2\omega_f$.

When the amplitude is not small,
the oscillations are also not simple and include various modes.
In order to separate these modes, 
we use the strength functions $S_{B,F}(\omega)$, 
which is defined 
as the Fourier transform of $Q_{B,F}$:
\begin{equation}
     S_{B,F}(\omega) = 
          \int^{t_f}_{t_i} d\tau\, Q_{B,F} (\tau) \sin{\omega \tau} 
\label{stFn}
\end{equation}
where we use the oddness $Q_{B,F}(\tau) =-Q_{B,F}(-\tau)$.
It is expected from the initial conditions in Eqs.~(\ref{bin}) and (\ref{fin}) 
because they lead to 
$Q_{B,F}= 0$ and $d Q_{B,F} /d \tau \neq 0$ at $\tau=0$.
For the time-integration in Eq.~(\ref{stFn}), 
we use the interval $0< \tau <200$ 
unless otherwise noted. 

Finally, 
we mention how the time-dependence of the collective oscillations 
obtained in the time-evolution method should be compared 
with the strength function in RPA in comparison.
The initial boosted state corresponding to the initial condition 
in Eqs.~(\ref{bin}) and (\ref{fin}) is given by
\begin{equation}
     |\Phi(\tau=0) \rangle =
          e^{\frac{i}{2} \left\{ \kappa_B {\hat O}_B +\kappa_F {\hat O}_F \right\}}
          |\Phi_0 \rangle,
\label{gsRPA}
\end{equation}
where $|\Phi_0 \rangle$ is the ground state, 
${\hat O}_{B,F}$ are the quadrupole operators defined 
in Eqs.~(\ref{qpOB}) and (\ref{qpOF}), 
and $\kappa_{B,F}$ correspond to the boost parameters.
Expanding $|\Phi(\tau=0) \rangle$ with respect to $\kappa_{B,F}$,
Eq.~(\ref{gsRPA}) becomes
\begin{eqnarray}
     |\Phi(\tau=0) \rangle &\approx& |\Phi_0 \rangle +
          \frac{i}{2} \left\{ \kappa_B {\hat O}_B + \kappa_F {\hat O}_F \right\}
          |\Phi_0 \rangle
\nonumber \\
                  &=&  |\Phi_0 \rangle 
                   +\frac{i}{2} \sum_n \left\{ \kappa_B T_B(\omega_n)
                                              +\kappa_F T_F (\omega_n) \right\}
                        |\Phi_n \rangle, 
\end{eqnarray}
where the transition amplitudes $T_{B,F}(\omega_n)$ are given 
in Eqs.~(\ref{ABn}) and (\ref{AFn}).
Thus, using the strength functions $T_{B,F}$, 
which is obtained in RPA, 
the time-dependent functions $Q_{B,F}$ are represented by
\begin{eqnarray}
     Q_B(\tau) &=& \frac{1}{N_b R_0^2(B)} \sum_n T_B(\omega_n)
                   \left\{ \kappa_B T_B(\omega_n) 
                          +\kappa_F T_F(\omega_n) \right\} \sin(\omega_n \tau),
\label{zBrpa}\\
     Q_F (\tau) &=& \frac{1}{N_f R_0^2(F)} \sum_n T_F(\omega_n)
                    \left\{ \kappa_B T_B(\omega_n) 
                           +\kappa_F T_F(\omega_n) \right\} \sin(\omega_n \tau).
\label{zFrpa}
\end{eqnarray}
Also, in RPA, 
the strength functions become
\begin{eqnarray}
     S_B(\omega) &=& \frac{1}{2N_b} \sum_n T_B(\omega_n)
                     \left\{ \kappa_B T_B(\omega_n) 
                            +\kappa_F T_F(\omega_n) \right\} F_n(t_f, t_i), \\
     S_F (\tau) &=& \frac{1}{2N_f} \sum_n T_F(\omega_n)
                    \left\{ \kappa_B T_B(\omega_n) 
                           +\kappa_F T_F(\omega_n) \right\} F_n(t_f, t_i),
\label{stFRPA}
\end{eqnarray}
where
\begin{equation}
     F_n(t_f, t_i) =\frac{ \sin{(\omega-\omega_n) t_f} -
                           \sin{(\omega-\omega_n) t_i}
                         }{2(\omega-\omega_n)}
                   -\frac{ \sin{(\omega+\omega_n) t_f} -
                           \sin{(\omega+\omega_n) t_i}
                         }{2(\omega+\omega_n)}.
\end{equation}

In the case of extremely small amplitude, 
the TDGP+Vlasov (time-evolution) approach and RPA should give the same results 
for the corresponding boost parameters: 
$\kappa_B = \lambda/\sqrt{2}$ and $\kappa_F = \lambda_F$;
it has been really confirmed in the case of dipole oscillations \cite{tbDPL}.
However, in the present calculation where the amplitude is not so small,  
the TDHF and Vlasov approach
does not give the same results with those in RPA;
especially, in the later stage of time-evolution.
Thus, we use the corresponding values of $\kappa_{B(F)}$ in RPA 
as to reproduce the oscillations in early stage of time-evolution 
in the TDGP and Vlasov approach.

\subsection{Quadrupole Oscillations in  $^{170}$Yb$-^{171}$Yb}
\label{QP1}

\begin{figure}[ht]
\hspace*{0cm}
{\includegraphics[scale=0.5]{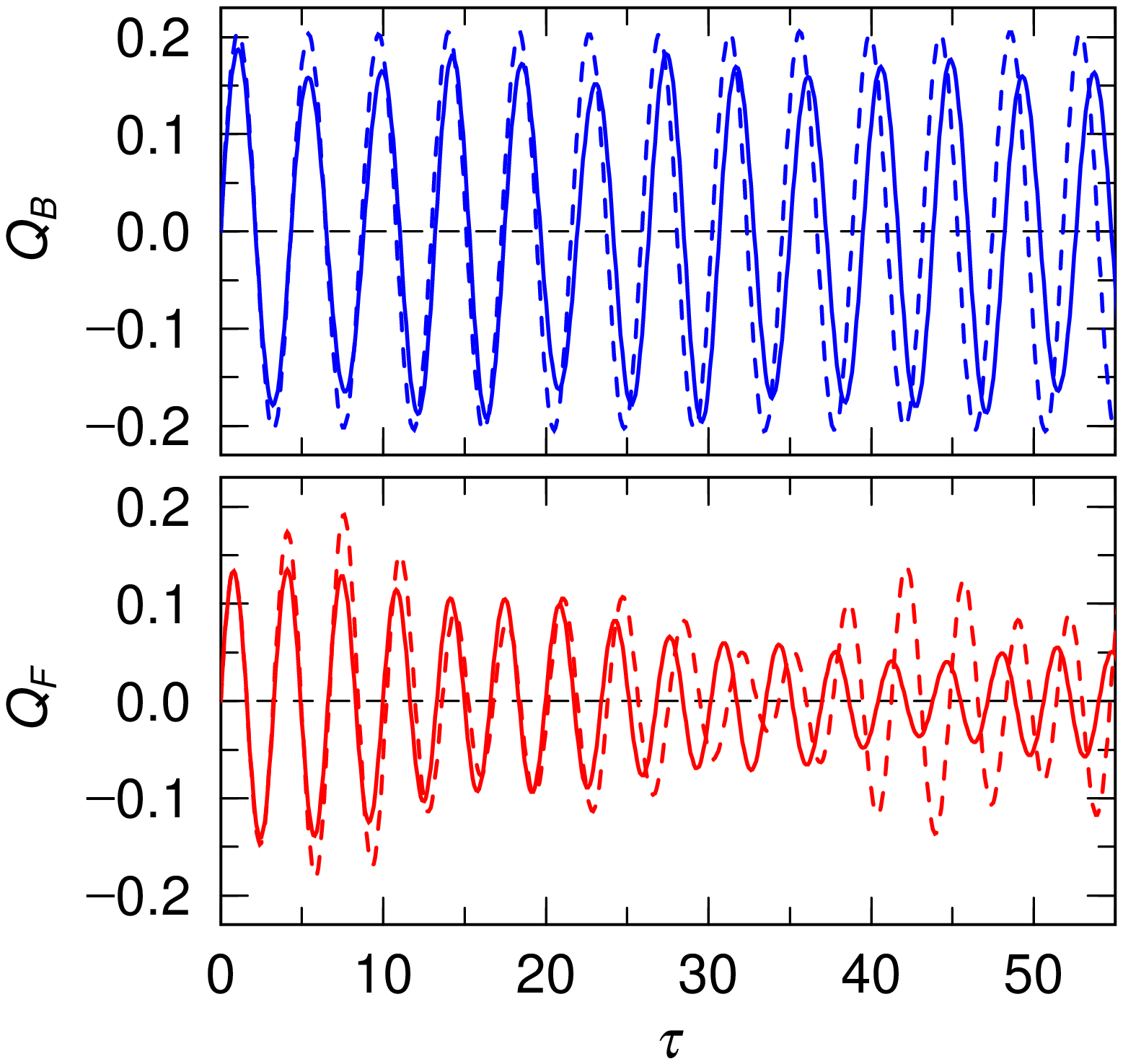}}
\caption{\small
Time evolution of the quadrupole oscillations 
in the $^{170}$Yb$-^{171}$Yb mixture 
with the in-phase initial condition
$\lambda_B = \lambda_F = 0.1$.
The upper and lower panels are 
$Q_{B,F}$ for the bosons and fermions,
and the solid and dashed lines represent the results of TDGP~+~Vlasov
and RPA calculations.}
\label{qYb1in}
\end{figure}

\begin{figure}[ht]
{\includegraphics[scale=0.5]{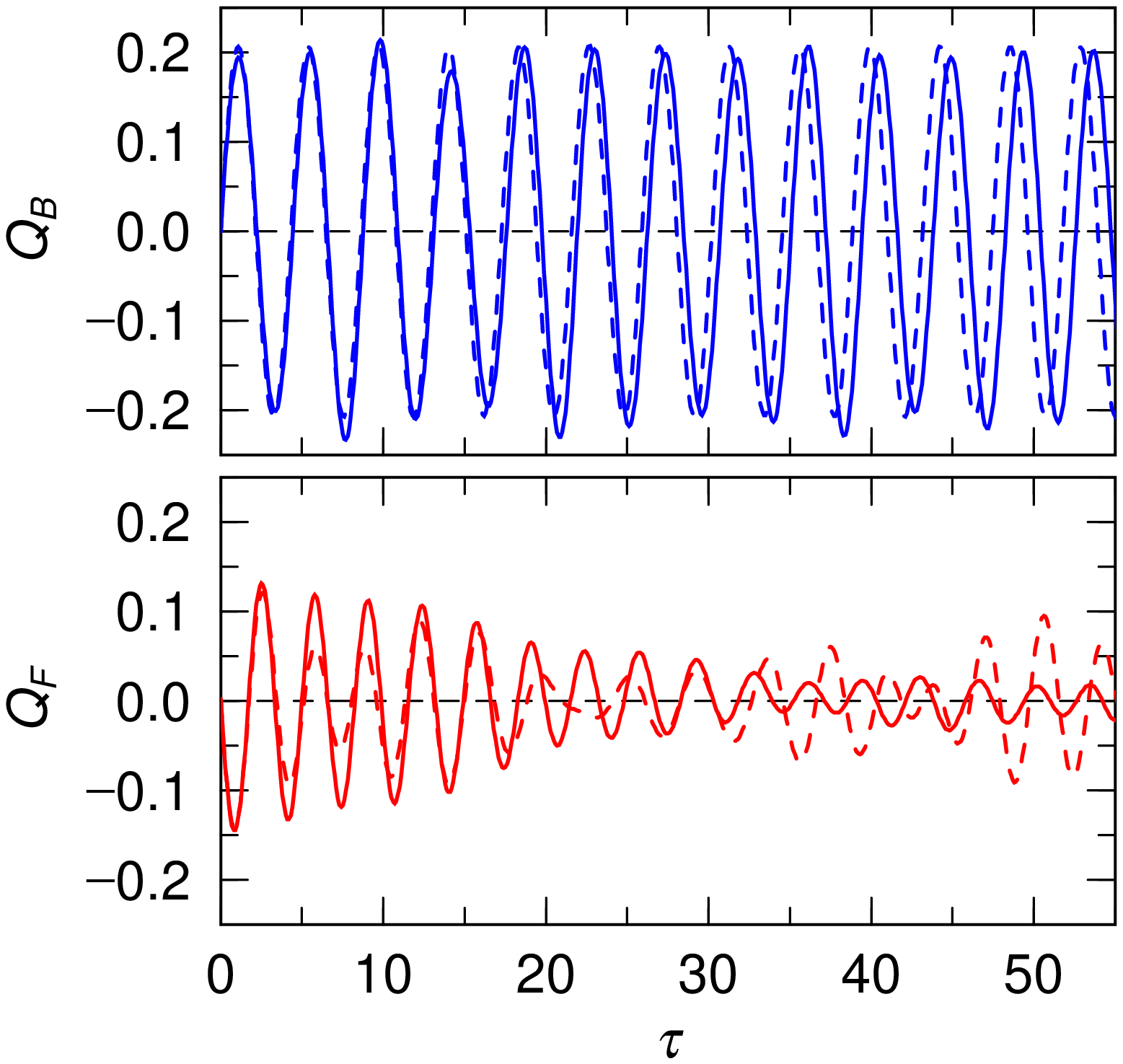}}
\caption{\small
Time evolution of the quadrupole oscillations 
in the $^{170}$Yb$-^{171}$Yb mixture 
with the out-of-phase initial condition
$\lambda_B = - \lambda_F = 0.1$.
The panels and the lines are the same as in Fig. \ref{qYb1in}.}
\label{qYb1ou}
\end{figure}

First, we discuss the quadrupole oscillation in \Ybfst system, 
where the boson-fermion interaction is weekly repulsive.
In Fig.~\ref{qYb1in}, 
we show the time-dependences of the 
$Q_{B}$ (the upper panel) and $Q_{F}$ (the lower panel) 
for the initial condition $\lambda_B=\lambda_F=0.1$,
which corresponds to the ``in-phase'' boson-fermion oscillation.  
The dashed and solid lines represent the results 
of the TDGP+Vlasov and RPA calculations, respectively.
Furthermore, we also show the results 
for the initial condition of the ``out-of-phase'' boson-fermion oscillation: 
$\lambda_B=-\lambda_F=0.1$ in Fig.~\ref{qYb1ou}.  

In both cases,
the results of the TDGP+Vlasov calculation agree well with those in RPA 
in the early stage of time-evolution ($\tau \lesssim 20$).
In RPA, 
the fermion oscillations show clear beating;
the amplitudes gradually decrease around $\tau \approx 35$ and 
become larger after that.
In the TDGP+Vlasov approach, 
the amplitudes are also damped but does not show clear beat phenomena.

\begin{figure}[ht]
\includegraphics[scale=0.5,angle=270]{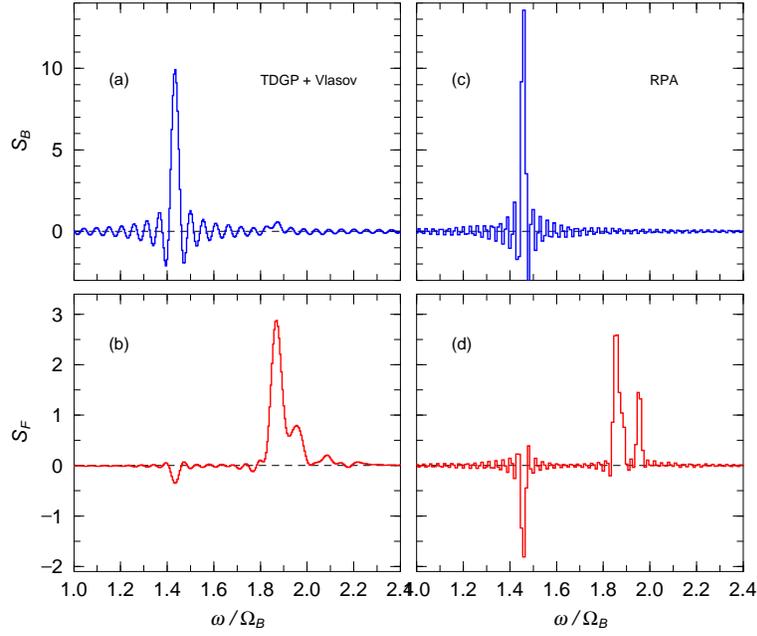}
\caption
{\small 
The strength functions of the quadrupole oscillations 
with the in-phase initial condition 
in {\Ybfst} mixture (corresponding to Fig.~\ref{qYb1in});
the boson and fermion oscillations in TDGP+Vlasov calculation (a and b), 
and in RPA calculation (c and d).}
\label{StYb1in}
\end{figure}

\begin{figure}[ht]
\vspace*{-0.5cm}
\includegraphics[scale=0.5,angle=270]{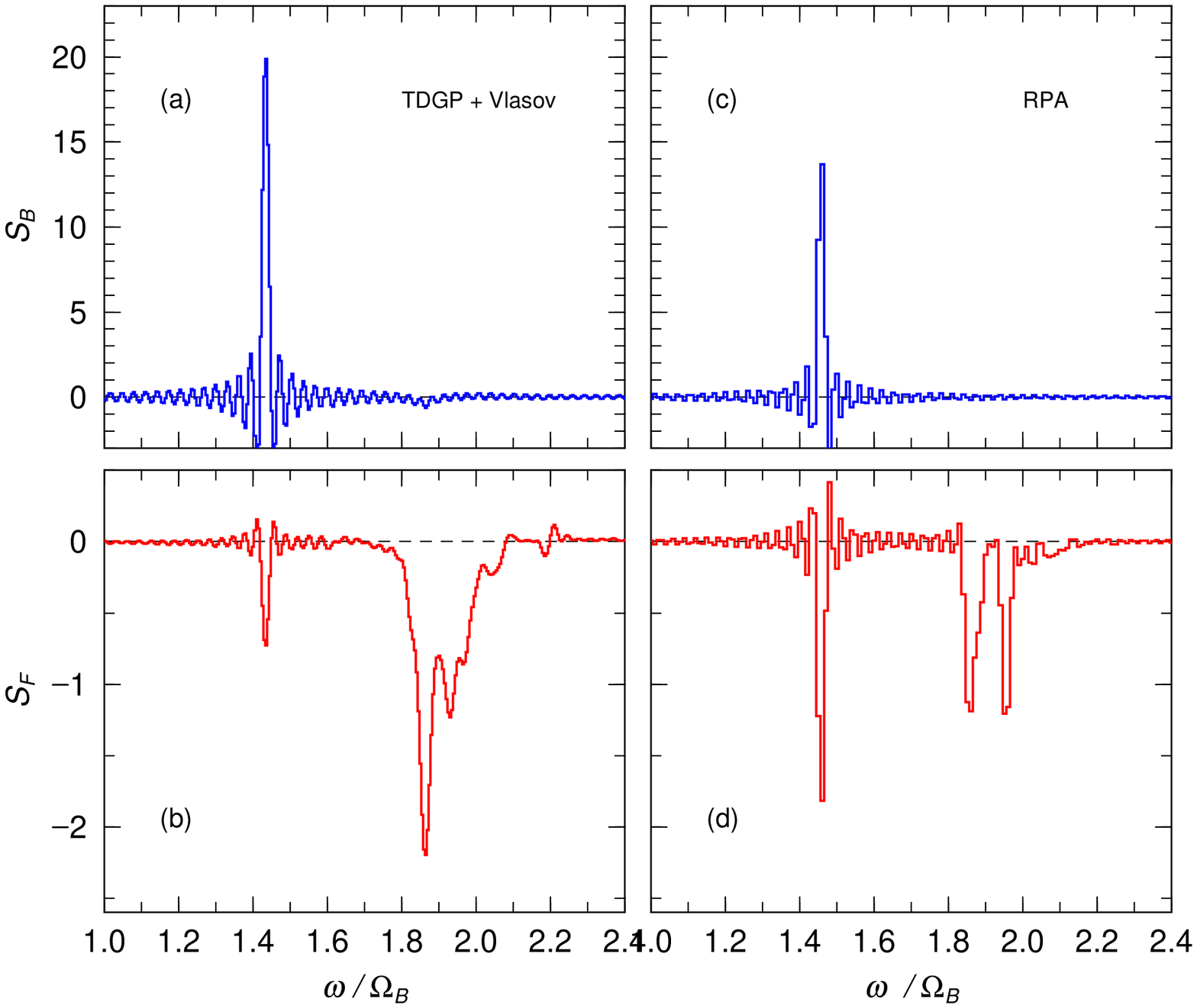}
\caption
{\small 
The strength functions of the quadrupole oscillations 
with the out-of-phase initial condition 
in {\Ybfst} mixture (corresponding to Fig.~\ref{qYb1ou});
The panels and the lines are the same with Fig. \ref{StYb1in}.
}
\label{StYb1ou}
\end{figure}

Next we calculate the strength functions $S_{B,F}$ for these oscillations 
using Eqs.~(\ref{stFn}) with integrating
over the range of $0 \le \tau \le 200$.
Figs.~\ref{StYb1in} and \ref{StYb1ou} show the $S_{B,F}$
for the oscillations with the in-phase and out-of-phase initial conditions; 
the boson and fermion strength functions in TDGP+Vlasov calculation
are in the upper left (a) and  the lower left (b) panels,
and those in  RPA are in the upper right (c) 
and the lower right (d) panels.

In these figures,
we find that the boson strength functions $S_B$ have only one sharp peak, 
which is consistent with the monotonous behavior of $Q_B(\tau)$ 
in Figs.~\ref{qYb1in} and \ref{qYb1ou}.
In contrast, the fermion strength functions $S_F$ has one negative peak 
at $\omega =\omega_Q^b$ and two peaks at $\omega \approx1,87$ and  
$\approx 1.96$; these two peaks are positive 
in the in-phase initial condition 
and negative in the out-of-phase initial condition. 

These behaviors are very similar to those in the dipole oscillations
\cite{tbDPL}.
The peaks at $\omega \approx \omega_Q^b$
and at $\omega \approx 1.87$
correspond to the boson-forced oscillation (mode-1).
and the inside-fermion oscillation (mode-3), respectively. 
In addition, 
we see a small peak at $\omega \approx 1.96$, 
which correspond to the outside-fermion oscillation (mode-2).
In TDGP+ Vlasov calculations, 
these peaks have broader widths than in RPA, 
so that the strengths of the outside-fermion oscillation (mode-2) 
in TDGP+Vlasov calculation are not so clear
(Figs.~\ref{qYb1in}b and \ref{qYb1ou}b).

\begin{figure}[ht]
{\includegraphics[scale=0.5]{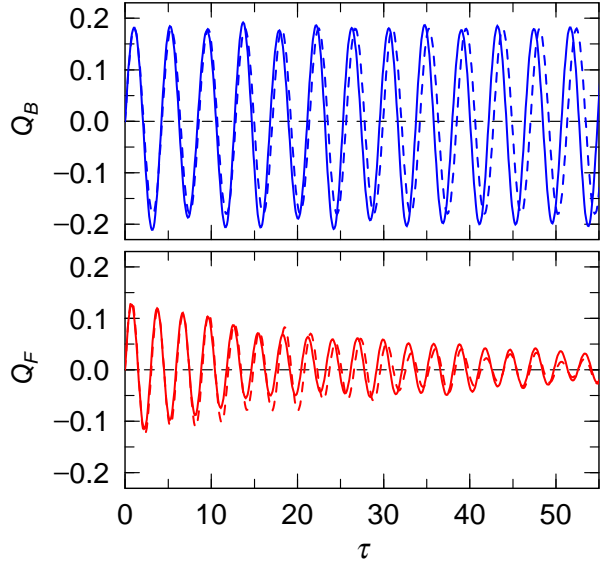}}
\caption{\small
Time evolution of the quadrupole oscillations 
with the in-phase initial condition
in the BF mixture with $h_{BF} = -1.9680$;
other parameters are the same as in $^{170}$Yb$-^{171}$Yb case.
The panels and the lines are the same as in Fig. \ref{StYb1in}.} 
\label{qYb1Min}
\end{figure}

\begin{figure}[ht]
\includegraphics[scale=0.5,angle=270]{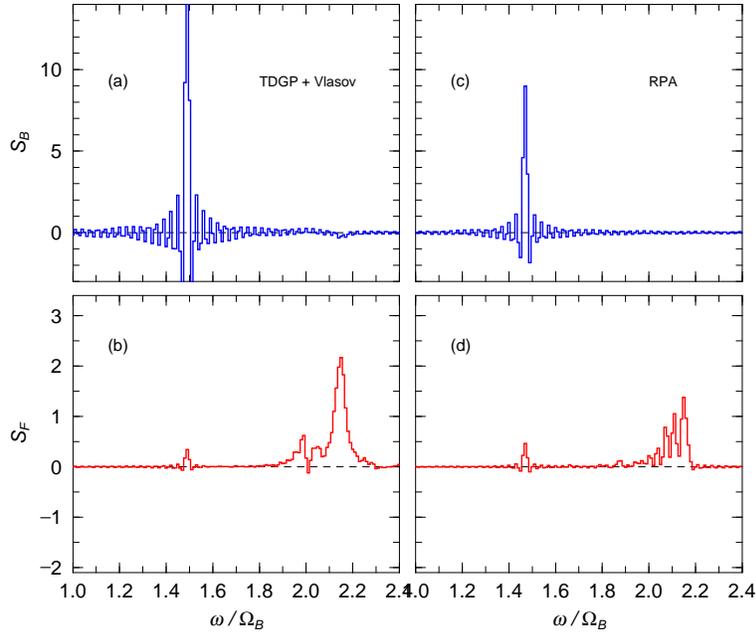}
\caption
{\small 
The strength functions of the quadrupole oscillations
corresponding to Fig.~\ref{qYb1Min};
other parameters are the same as in $^{170}$Yb$-^{171}$Yb case.
The panels and the lines are the same as in Fig. \ref{StYb1in}.} 
\label{StYb1Min}
\end{figure}

In order to examine the sign-dependence 
of the boson-fermion coupling constant $h_{BF}$, 
we perform an additional simulation 
with the attractive boson-fermion coupling constant 
with $h_{BF} = -1.9680$ 
(and the same values for the other parameters as in \Ybfst).
In Figs.~\ref{qYb1Min} and \ref{StYb1Min},
we show the time-evolutions $Q_{B,F}$and the strength functions $S_{B,F}$ 
with the in-phase initial condition.
The TDGP+Vlasov and RPA calculations give almost the same results. 
In this case, the fermion oscillation shows no beats 
in both approaches (Fig.~\ref{qYb1Min}, bottom).
In the strength functions (Fig.~\ref{StYb1Min}),
the peak of the fermion oscillation at $\omega=\omega_Q^b$ becomes positive, 
and its heights becomes smaller than that in Fig.~\ref{StYb1in}.
We also see a peak at $\omega \approx 1.99$ in $S_F$ 
in the TDGP+Vlasov calculation  (Fig.~\ref{StYb1Min}b), 
which corresponds to the outside-fermion oscillation (mode-2), 
but no clear peaks are found around this frequency in the RPA calculation.
In the case of attractive boson-fermion interaction, 
more fermions populate in the boson-distributed region, 
and it makes the outside-fermion oscillation (mode-2) strength small.
Thus, the strength of the mode-2 becomes very small
in the small-amplitude oscillations described in RPA.
As the amplitude becomes larger, however, 
more fermions move to the outside 
from the boson-distributed region, 
and the strength of the mode-2 becomes larger in TDGP+Vlasov approach.

\subsection{Quadrupole Oscillations in  $^{170}$Yb$-^{173}$Yb}
\label{QP}

In this subsection,
we discuss the quadrupole oscillations in the \Ybsnd system, 
where the boson-fermion interaction is strongly attractive.

\begin{figure}[ht]
{\includegraphics[scale=0.5]{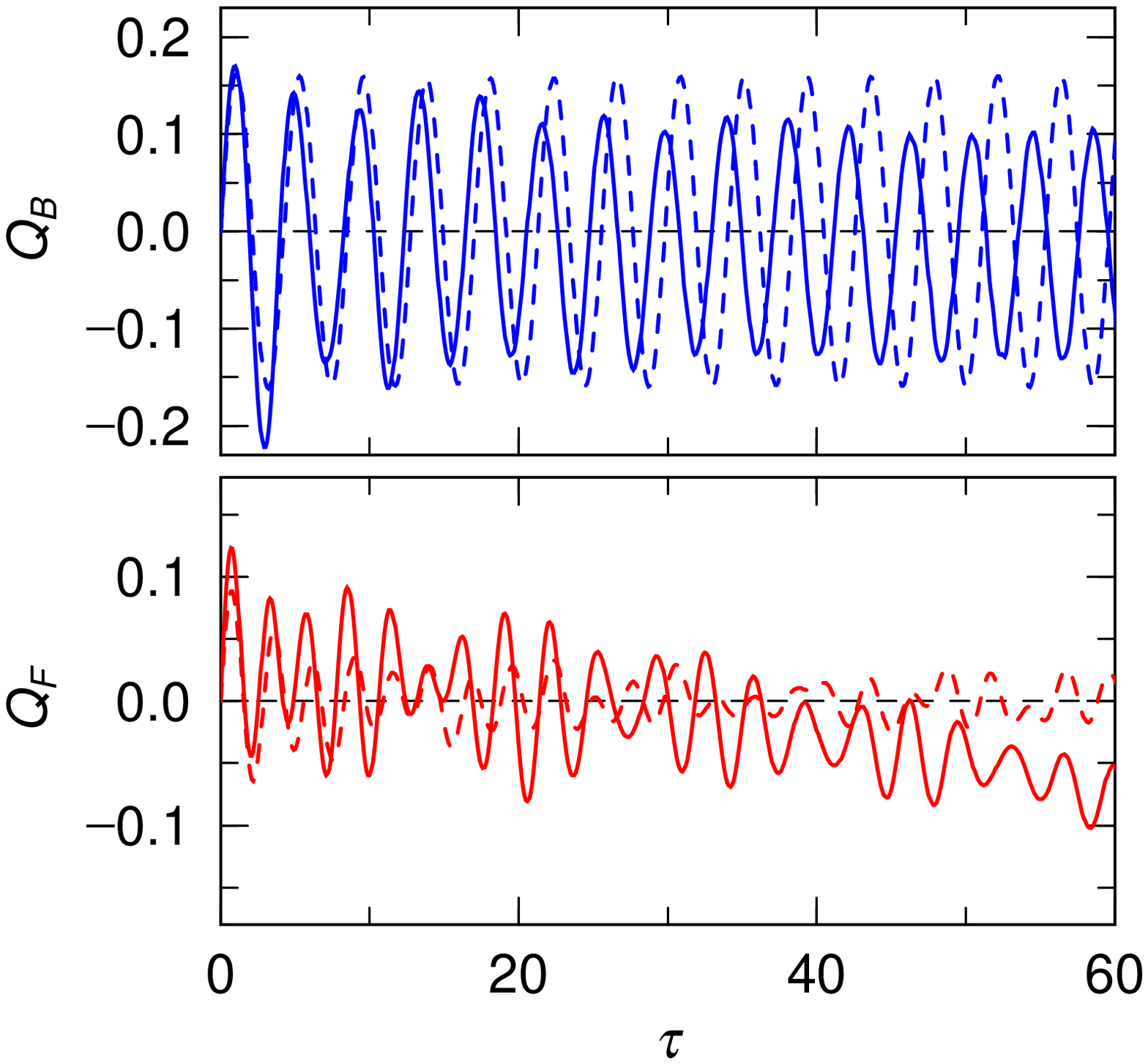}}
\caption{\small
Time evolution of the quadrupole oscillations 
in the $^{170}$Yb$-^{173}$Yb mixture 
with the in-phase initial condition
$\lambda_B = \lambda_F = 0.1$.
The upper and lower panels are 
$Q_{B,F}$ for the bosons and fermions,
and the solid and dashed lines represent the results of TDGP~+~Vlasov
and RPA calculations.}
\label{qYb2in}
\end{figure}

\begin{figure}[ht]
\vspace*{-0.5cm}
{\includegraphics[scale=0.5]{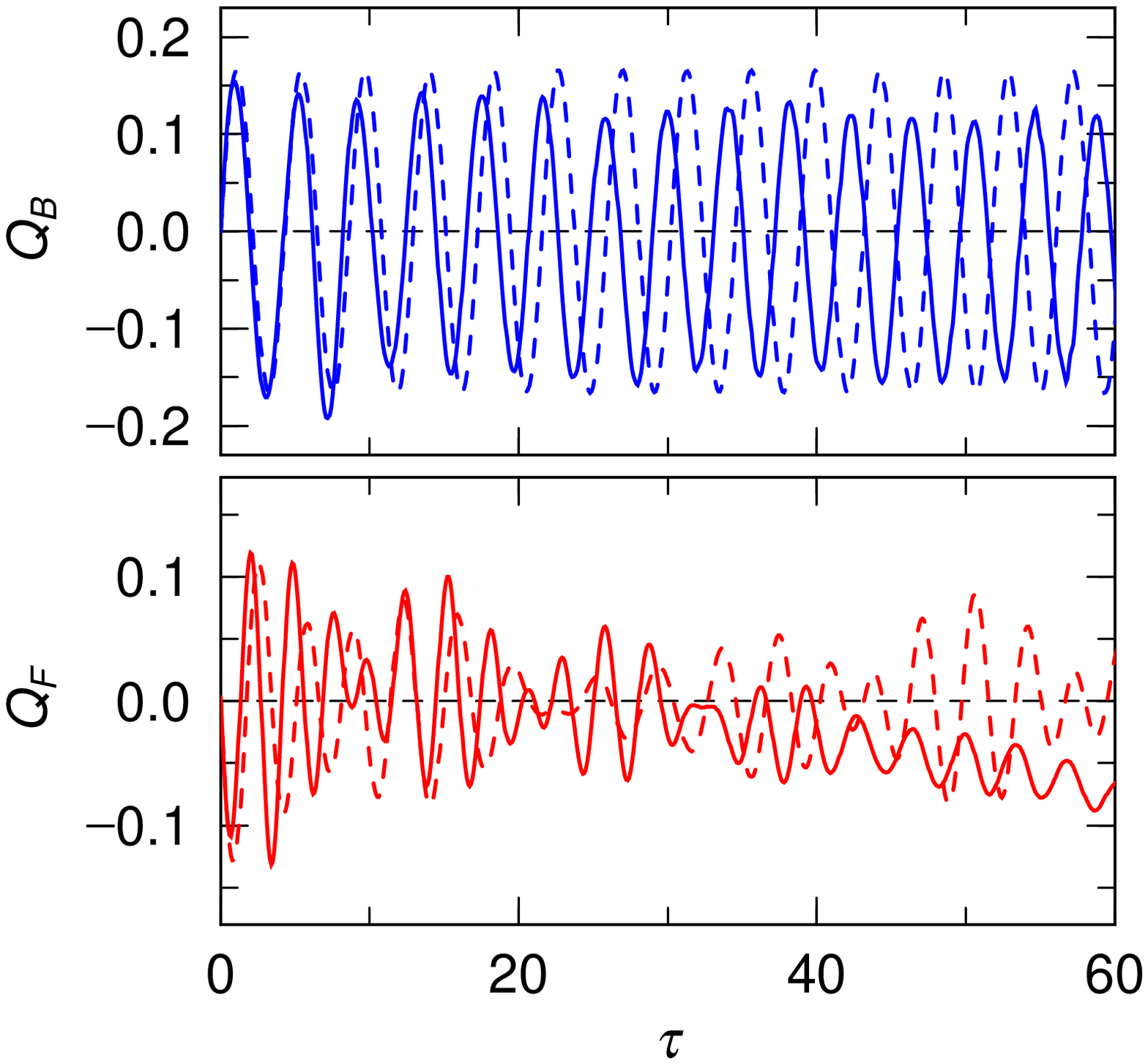}}
\caption{\small
Time evolution of the quadrupole oscillations 
in the $^{170}$Yb$-^{173}$Yb mixture 
with the out-of-phase initial condition
$\lambda_B = - \lambda_F = 0.1$.
The panels and the lines are the same as in Fig. \ref{qYb2in}.}
\label{qYb2ou}
\end{figure}

In Figs.~\ref{qYb2in} and \ref{qYb2ou},
we show the time-dependence of the $Q_{B,F}$ in \Ybsnd system 
for two kinds of initial conditions, $\lambda_B=\lambda_F=0.1$ (in-phase)
and $\lambda_B=-\lambda_F=0.1$ (out-of-phase), respectively.

Figs.~\ref{qYb2in} and \ref{qYb2ou} clearly show that 
the TDGP+Vlasov calculation gives very different results from RPA in this BF mixture.
The boson oscillations in the TDGP+Vlasov calculation have slight damping, 
and shorter periods than those in RPA.
Furthermore, $Q_F$ decrease in the TDGP+Vlasov calculation and
become negative when $\tau \gtrsim 40$ (bottom panels).
In comparison with RPA, 
the TDGP+Vlasov calculation shows slightly shorter periods in the bose oscillations
and exhibit slight dampings (top panels).

\begin{figure}[ht]
\includegraphics[scale=0.8]{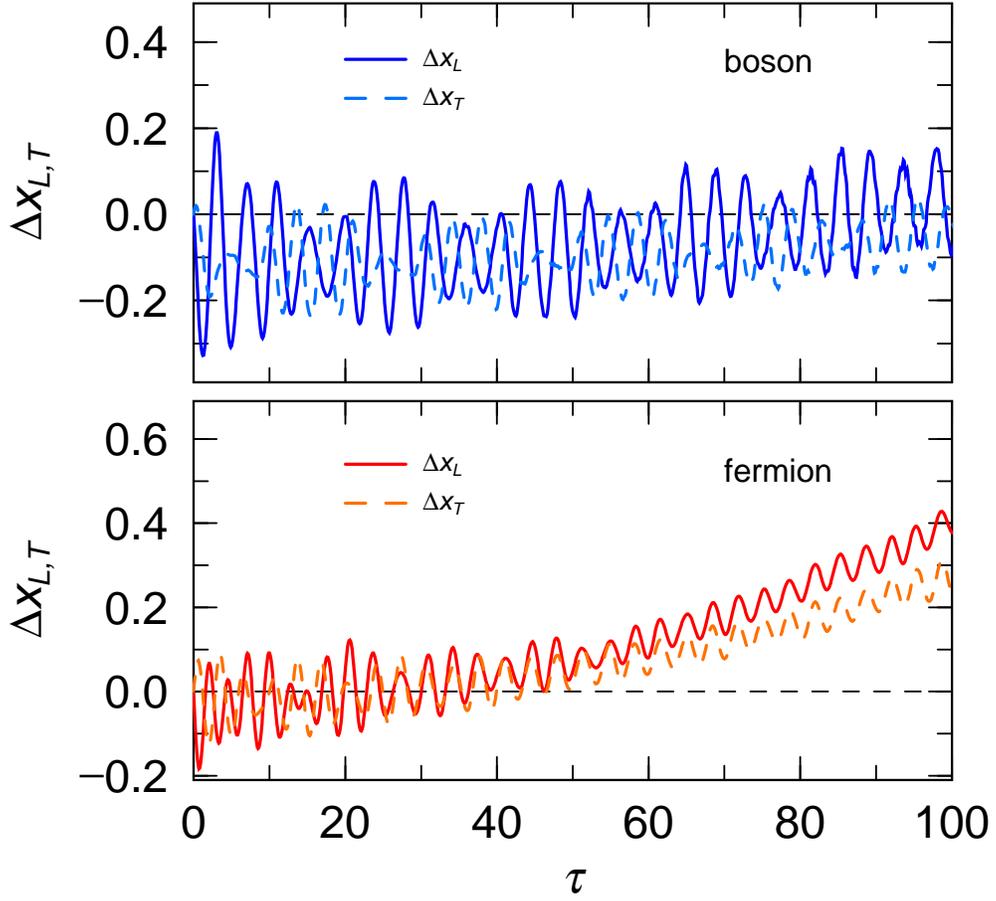}
\caption{\small 
Time evolution of $R_z$ (solid lines) and $R_T$ (dotted lines) 
for the quadrupole oscillation 
in \Ybsnd mixture 
with the in-phase initial condition
(corresponding to Fig.~\ref{qYb2in});
the upper and lower panels are for the bosons and fermions.}
\label{BrQ2in}
\end{figure}

In order to examine these results further,
we calculate the deformation shifts in the longitudinal and transversal directions: 
$\Delta x_L =  \sqrt{3} R_L / R_0 - 1$ and 
$\Delta x_T =  \sqrt{3/2} R_T / R_0 - 1$, 
where $R_{0,L,T}$ are the root-mean-square radii of the ground and the oscillation states, 
which are defined in Eqs.~(\ref{RT}) and (\ref{RL}). 
The time-dependences of $\Delta x_{L,T}$ are plotted in Fig.~\ref{BrQ2in} 
in the case of the in-phase initial condition. 
In these figures, 
we can find the increase of the fermion radii  
in both directions;
namely the expansion of the fermion gas occurs in the oscillation process.
We can also find $R_L > R_T$ for the fermion oscillations in this expanding process;
from the definition of $Q_F$ in (\ref{QBF}), 
it is just the origin of $Q_F < 0$ (Figs.~\ref{qYb2in} and \ref{qYb2ou}, bottom).
This expansion can be understood by the fermion overflow from the boson-distributed region;
in the ground state, fermions populate largely in the boson-distributed region 
because of the attractive BF interaction, but, in the excited sate, 
some of them are pushed into the outside in the course of oscillation.
Such expansion has already been discussed theoretically 
in the monopole oscillation \cite{monoEX}. 

\begin{figure}[ht]
\includegraphics[scale=0.5,angle=270]{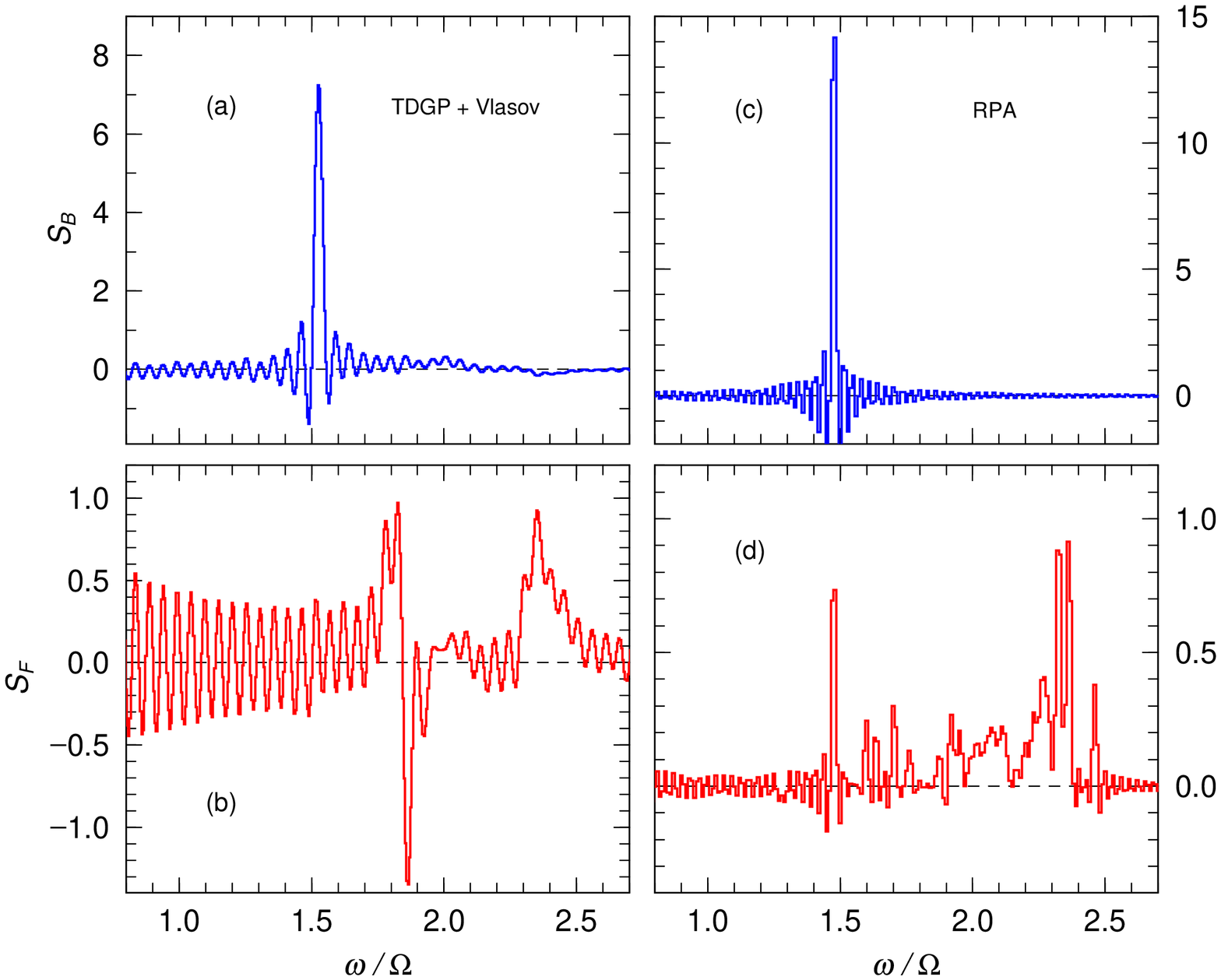}
\caption
{\small 
The strength functions of the quadrupole oscillations 
with the in-phase initial condition 
in {\Ybsnd} mixture (corresponding to Fig.~\ref{qYb2in});
the panels and lines are the same as in Fig.~\ref{StYb1in}.}
\label{StYb2in}
\end{figure}

\begin{figure}[ht]
\vspace*{-1.5cm}
\includegraphics[scale=0.5,angle=270]{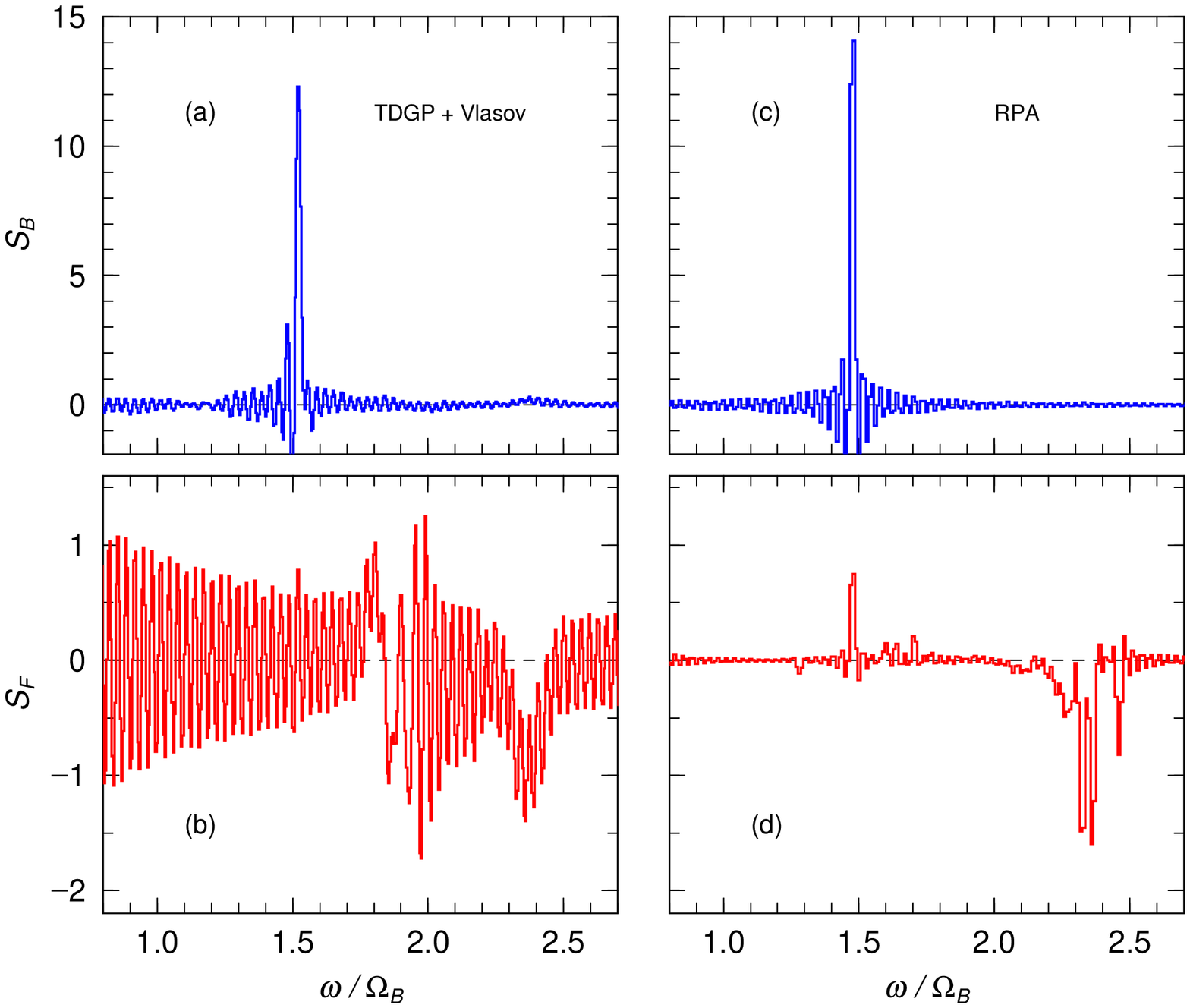}
\caption
{\small 
The strength functions of the quadrupole oscillations 
with the out-of-phase initial condition 
in {\Ybsnd} mixture (corresponding to Fig.~\ref{qYb2ou});
The panels and the lines are the same with Fig. \ref{StYb1in}.
}
\label{StYb2ou}
\end{figure}

The strength functions $S_{B,F}$ corresponding to these oscillations are shown 
in Figs.~\ref{StYb2in} (in-phase) and \ref{StYb2ou} (out-of-phase), 
respectively.
In RPA calculations (Figs.~\ref{StYb2in}c,d and \ref{StYb2ou}c,d), 
the $S_{B,F}$ have a clear peak at $\omega \approx 1.4$
and a broad peak around $\omega \approx 2.3$,
which correspond to the boson-forced (mode-1) and the inside-fermion (mode-3) oscillations. 
No peaks exist, which correspond to the outside oscillation (mode-2).
It is because, in the ground state,
the attractive interaction makes a large part of fermions populate 
in the boson-distributed region 
so that the fermions that exist outside region 
gives very small contribution in $S_{B,F}$.

The TDGP+Vlasov calculation  
gives similar results for the $S_B$ with those in RPA
(Figs.~\ref{StYb2in}a and \ref{StYb2ou}a);
however, the $S_F$ are quite different
(Figs.~\ref{StYb2in}b and \ref{StYb2ou}b).
In the TDGP+Vlasov calculation, 
the $S_F$ have no clear peak at $\omega \approx 1.4$, 
which corresponds to the boson-forced oscillation (mode-1), 
but have a broad peak appears around $\omega \approx 2.3$.
Also the strength below $\omega \lesssim 1.4$ monotonously
increases as $\omega$ becomes smaller; 
these monotonous behaviors in the $S_F$ are also caused by the expansion of the fermion gases.

\begin{figure}[ht]
\vspace*{-0.5cm}
\includegraphics[scale=0.5]{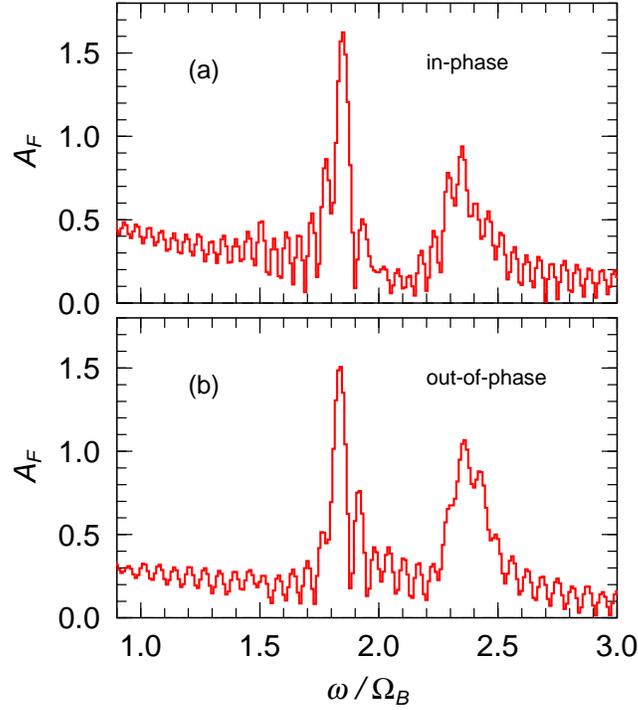}
\caption
{\small 
The fermion absolute strength functions of the quadrupole oscillations in \Ybsnd
gas with the in-phase and the out-of-phase initial conditions.
}
\label{AstQf}
\end{figure}

\begin{figure}[ht]
\vspace*{-0.5cm}
\includegraphics[scale=0.5]{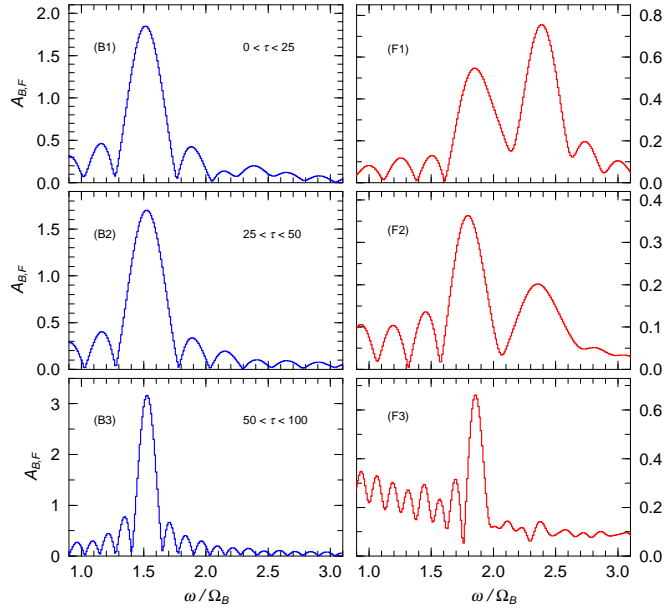}
\caption
{\small 
The absolute strength functions of the quadrupole oscillations in \Ybsnd mixtures 
with the in-phase initial conditions;
the left (B1,B2,B3) and right (F1,F2,F3) panels are 
for the boson and fermion oscillations.
The strength functions are deduced from the evolution in $0 < \tau < 25$ 
(the top panels),
in $25 < \tau < 50$ (the middle panels)
and in $50 < \tau < 100$ (the bottom panels), respectively.}
\label{stQT}
\end{figure}

Furthermore,
we see an unclear peak around $\omega \approx 1.85$ in the $S_F$ (Fig.~\ref{StYb2in}d).
As mentioned before, 
the $Q_F$ is assumed to be an odd function for the time parameter $\tau$, 
but the calculated strength function does not seem
to be the odd function around this peak.
In order to clarify the situation, 
we calculate the absolute strength function:
\begin{equation} 
A_{B,F}(\omega) = 
 \left| \int^{t_f}_{t_i} d \tau Q_{B,F} (\tau) e^{i\omega \tau} \right|.
\label{AstFn}
\end{equation}
The results are shown in Fig.~\ref{AstQf}a,b for the oscillations 
with the in-phase
and the out-of-phase initial conditions.
Two clear peaks are seen at $\omega = 1.85$ and 2.35 in both cases. 
To make clear the origin of these peaks, 
we divide the oscillation period into three regions
$0<\tau<25$ (F1, early stage), $25<\tau<50$  (F2, intermediate stage) 
and $50<\tau<100$ (F3, later stage), 
and evaluate the $A_F(\omega)$ in these stages.
As shown in Fig.~\ref{stQT}, 
the peak of the $S_F$  appears around $\omega=2.35$ 
in the early stage (panel F1), but disappear in the latter stage (panel F3).
In contrast, the peak at $\omega=1.85$ is seen in all stages.

\begin{figure}[ht]
\hspace*{-1.5cm}
\includegraphics[scale=0.5]{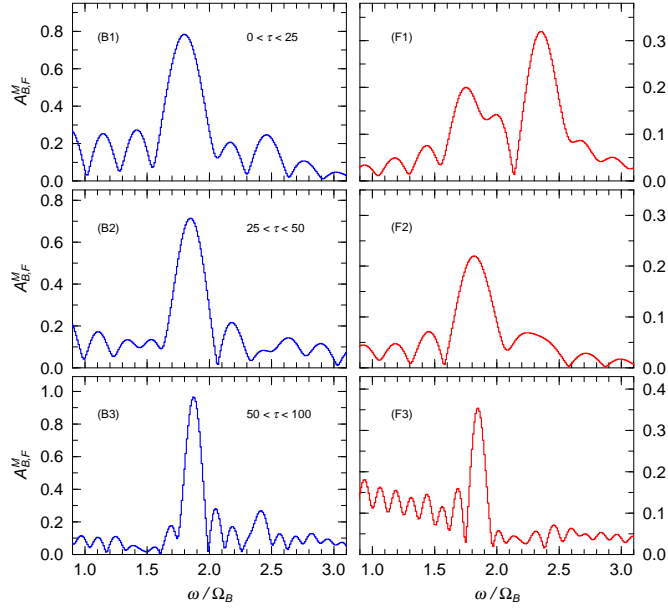}
\caption{
The absolute strength functions of the monopole oscillations in \Ybsnd mixtures 
with the in-phase initial conditions.
The panels are the same as in Fig.~\ref{stQT}.}
\label{stMT}
\end{figure}

Because of the large expansion of the fermion gas,
which is shown in Fig.~\ref{BrQ2in}, 
the quadrupole oscillations are supposed to couple 
with the monopole oscillation mode.  
For examination, 
we calculate the strength function of the
monopole oscillation mode: 
\begin{equation}
     A^M_{B,F} (\omega) = 
          \left|\int^{t_f}_{t_i} d\tau 
          \left[\frac{R(B,F)}{R_0(B,F)} - 1 \right] e^{i\omega \tau} \right|,
\label{AstM}
\end{equation}
which are plotted in Fig.~\ref{stMT} for the three stages: 
$0<\tau<25$ (B1,F1),
$25<\tau<50$ (B2,F2) and $50<\tau<100$ (B3,F3). 
The fermion strength function has a peak at $\omega \approx 2.35$
in early stage (F1) 
but it disappear in later stage (F2,F3);
on the other hand,  
the peak at $\omega \approx 1.85$ can be seen in all stages. 
Thus, we can find that the peak at $\omega \approx 1.85$ in the $S_F$ 
of the quadrupole oscillation corresponds to the monopole mode, 
which is excited through the expansion of the fermion gas; 
it explains the asymmetry of the $S_F$ around the peak.

From the above analysis,
we can obtain the following picture 
on the quadrupole oscillation in the \Ybsnd mixture:
1) The strongly-attractive boson-fermion interaction 
makes a large part of fermions populate 
in the boson-distributed region in the ground state.
2) In the quadrupole oscillation, 
the oscillating bosons causes the fermion overflow 
from this region
and it caused the expansion of the fermion gas. 
3) The initial conditions in the present calculation 
breaks the rotation symmetry, 
and,
in the large-amplitude oscillation,
the angular momentum becomes much larger than $2 \hbar$.
It can excite the monopole mode through the oscillation.
As a result, 
the oscillations include both the monopole and quadrupole modes.
4) In the course of the fermion gas expansion,
the intrinsic modes of the fermion quadrupole oscillation (mode-2 and -3) 
are damped and loses its strength in early stage;
instead, the monopole oscillation mode grows up.
5) In addition the expansion process of the fermion gas is slow and not
collective, and then
many small modes appear below $\omega \lesssim 1.5$.

%
%
\subsection{Quadrupole Oscillations in  $^{174}$Yb$-^{173}$Yb}
\label{QP3}
%
%

Finally,
we discuss the quadrupole oscillations in the \Ybthd mixture, 
where the boson-fermion interaction is strongly repulsive
and, different from other two mixtures, 
the fermion density distribution is surface-peaked.
 
When $h_{BF} > g_{BB}$, 
the effective potential for fermions has the minimum 
in the surface of the boson-distributed region, 
which produce the surface-peaked fermion-density distributions 
in the ground state; 
in the oscillation mode,  
the fermion gas is considered to oscillate 
in the neighborhood of this minimum.

\begin{figure}[ht]
\hspace*{0cm}
{\includegraphics[scale=0.6]{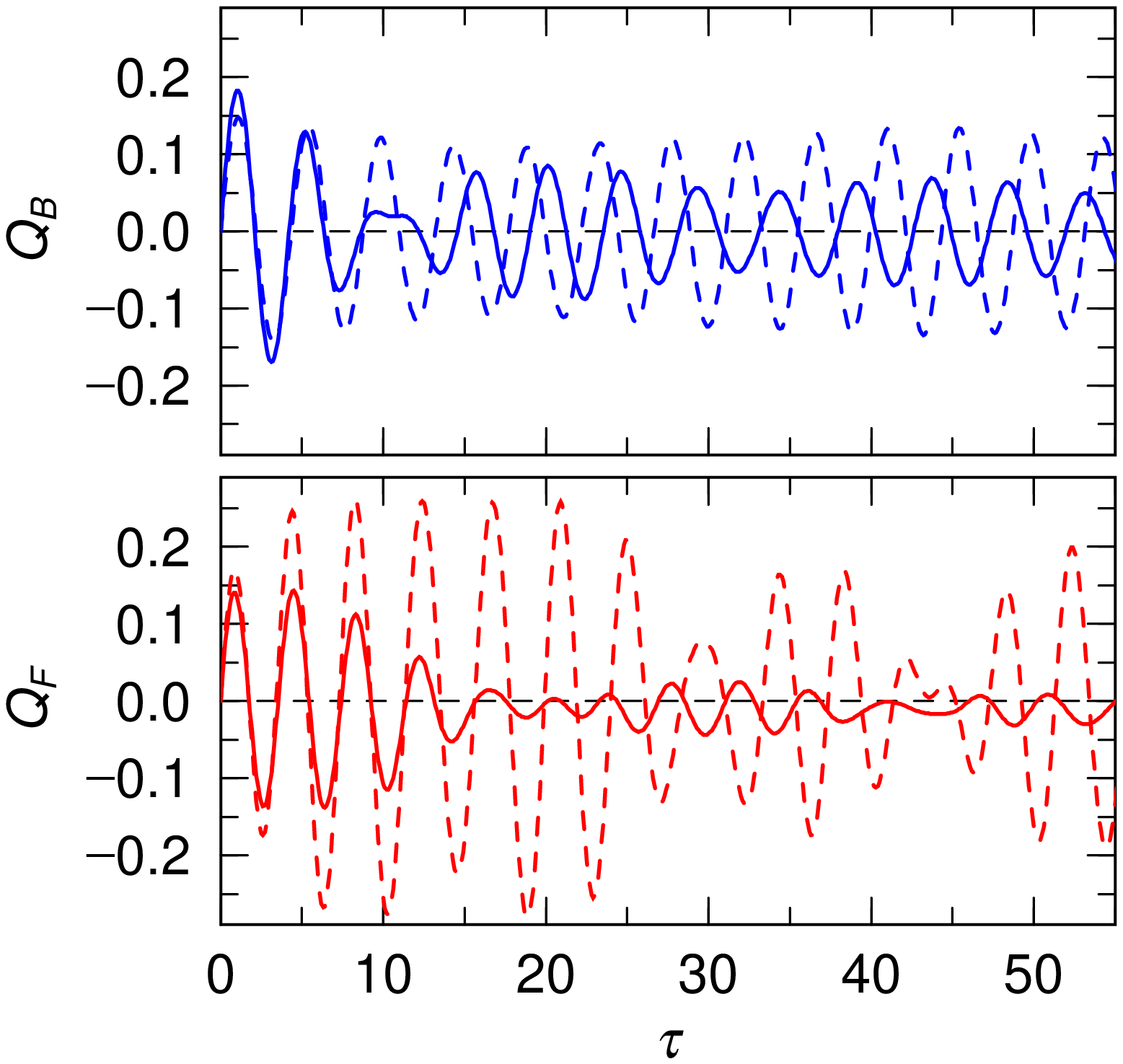}}
\caption{\small
Time evolution of the quadrupole oscillations 
in the \Ybthd mixture 
with the in-phase initial condition
$\lambda_B = \lambda_F = 0.1$.
The upper and lower panels are 
$Q_{B,F}$ for the bosons and fermions,
and the solid and dashed lines represent the results of TDGP~+~Vlasov
and RPA calculations.}
\label{qYb3in}
\end{figure}

\begin{figure}[ht]
{\includegraphics[scale=0.6]{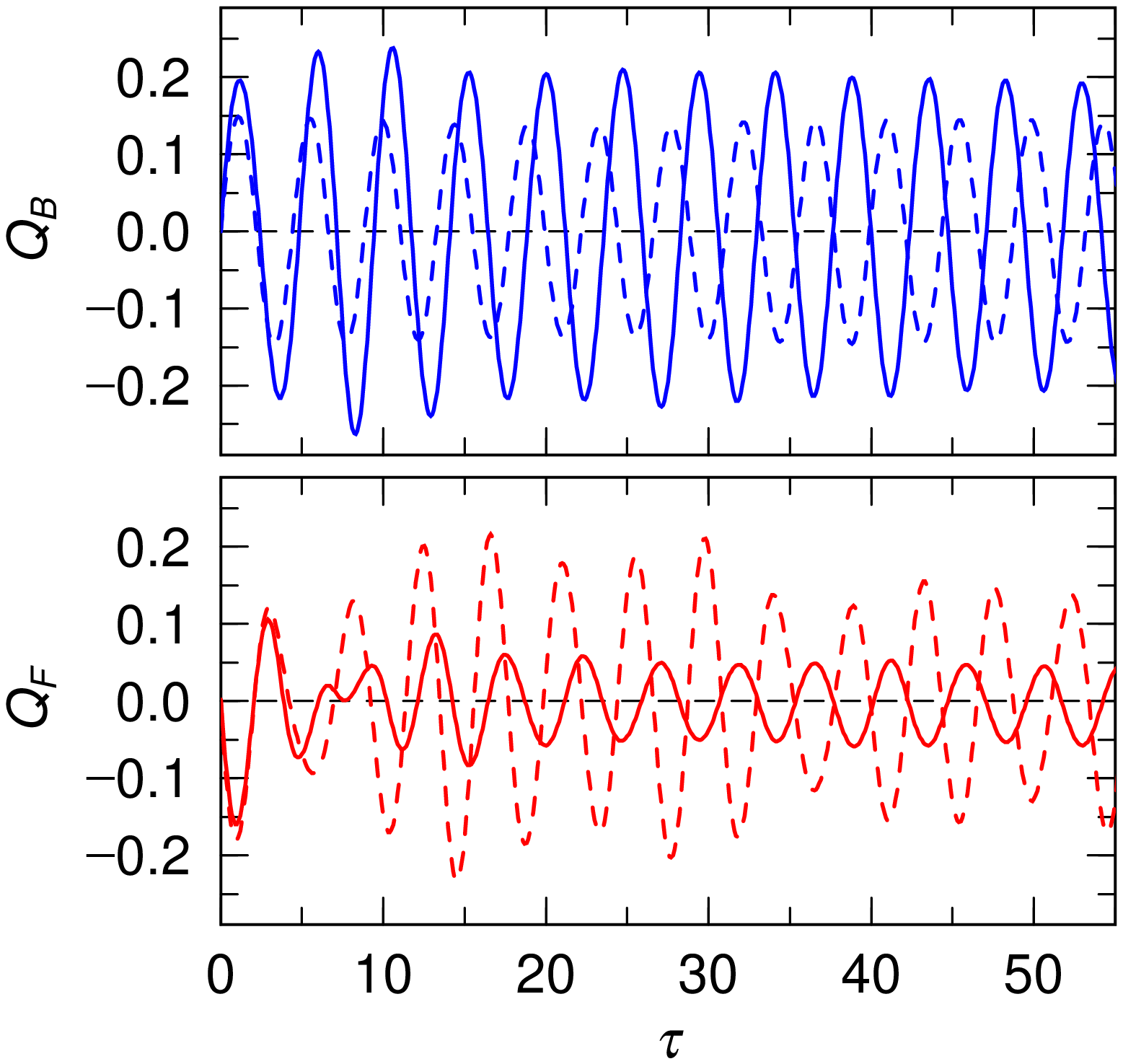}}
\caption{\small
Time evolution of the quadrupole oscillations 
in the \Ybthd mixture 
with the out-of-phase initial condition
$\lambda_B = - \lambda_F = 0.1$.
The panels and the lines are the same as in Fig. \ref{qYb2in}.}
\label{qYb3ou}
\end{figure}

Figs.~\ref{qYb3in} and \ref{qYb3ou} show the time-dependence 
of the $Q_{B,F}$ 
for the in-phase and out-of-phase initial conditions; 
the dashed and solid lines are 
for the TDGP+Vlasov and RPA calculations. 
These two calculations are consistent within one cycle of oscillation, 
but become very different after that. 
In the TDGP+Vlasov calculations, 
the amplitude of the fermion oscillation shows quick damping, 
and the boson oscillations exhibits beating and small damping
in the case of the in-phase initial condition (Fig.~\ref{qYb3ou}).

The similar damping phenomena have already been appeared  
in the monopole \cite{tomoBF} and dipole oscillations \cite{tbDPL};
however, in these oscillations, 
the boson oscillations are stable 
and simply affects the fermion motions 
through the strong repulsive BF interactions,
for quick dampings of the fermion oscillations 
decrease the strength of their intrinsic mode soon.

\begin{figure}[ht]
{\includegraphics[scale=0.6]{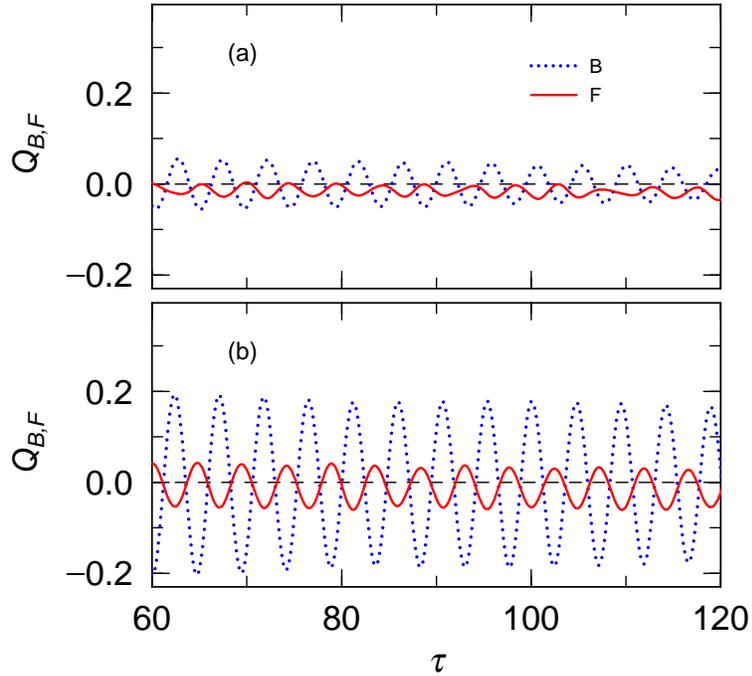}}
\caption{\small
Time evolution of the quadrupole oscillations in the \Ybthd mixture
with the in-phase (a) and the out-of-phase (b) initial conditions 
in the time-interval $60 < \tau < 120$;
the dashed and solid lines are for the boson and fermion
oscillations, respectively. }
\label{qYb3lat}
\end{figure}

In Fig.~\ref{qYb3lat}, 
we show those oscillations in later stage
$60 < \tau < 120$ 
in the cases of the in-phase (a)
and the out-of-phase  (b) initial conditions. 
We see that the oscillations keep 
the almost constant amplitudes  in this stage, 
and the bosons and fermions oscillate in out-of-phase 
with the same periods of oscillations.
These results shows that
the intrinsic fermion mode loses the strength in early stage, 
and only the forced-oscillation mode survives for a long time.

This mixture prefers the BF out-of-phase oscillation; 
even if the oscillation starts with the in-phase oscillation,
the relative phase shift occurs and 
the oscillation becomes the out-of-phase. 
At the time of the phase change, 
the oscillation accompanies large diffusion.
Thus, 
the $Q_F$ shows strong damping, 
and the $Q_B$ also shows small damping 
and beating throughout the process.

\begin{figure}[ht]
\hspace*{-1.0cm}
\includegraphics[scale=0.5,angle=270]{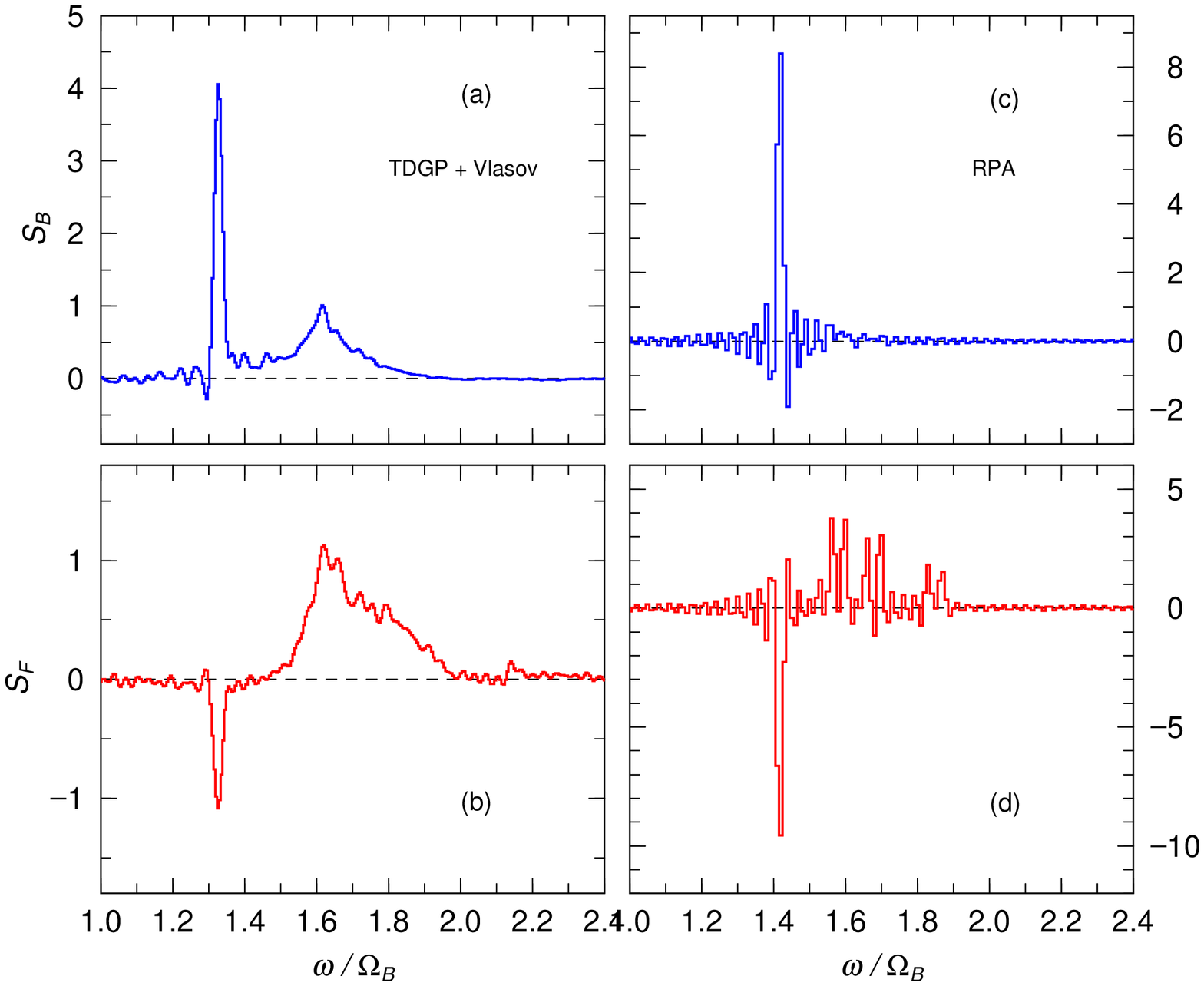}
\caption
{\small 
The strength functions of the quadrupole oscillations 
with the in-phase initial condition 
in \Ybthd mixture (corresponding to Fig.~\ref{qYb3in});
the panels and lines are the same as in Fig.~\ref{StYb1in}.}
\label{StYb3in}
\end{figure}

\begin{figure}[ht]
\hspace*{-1.0cm}
\includegraphics[scale=0.5,angle=270]{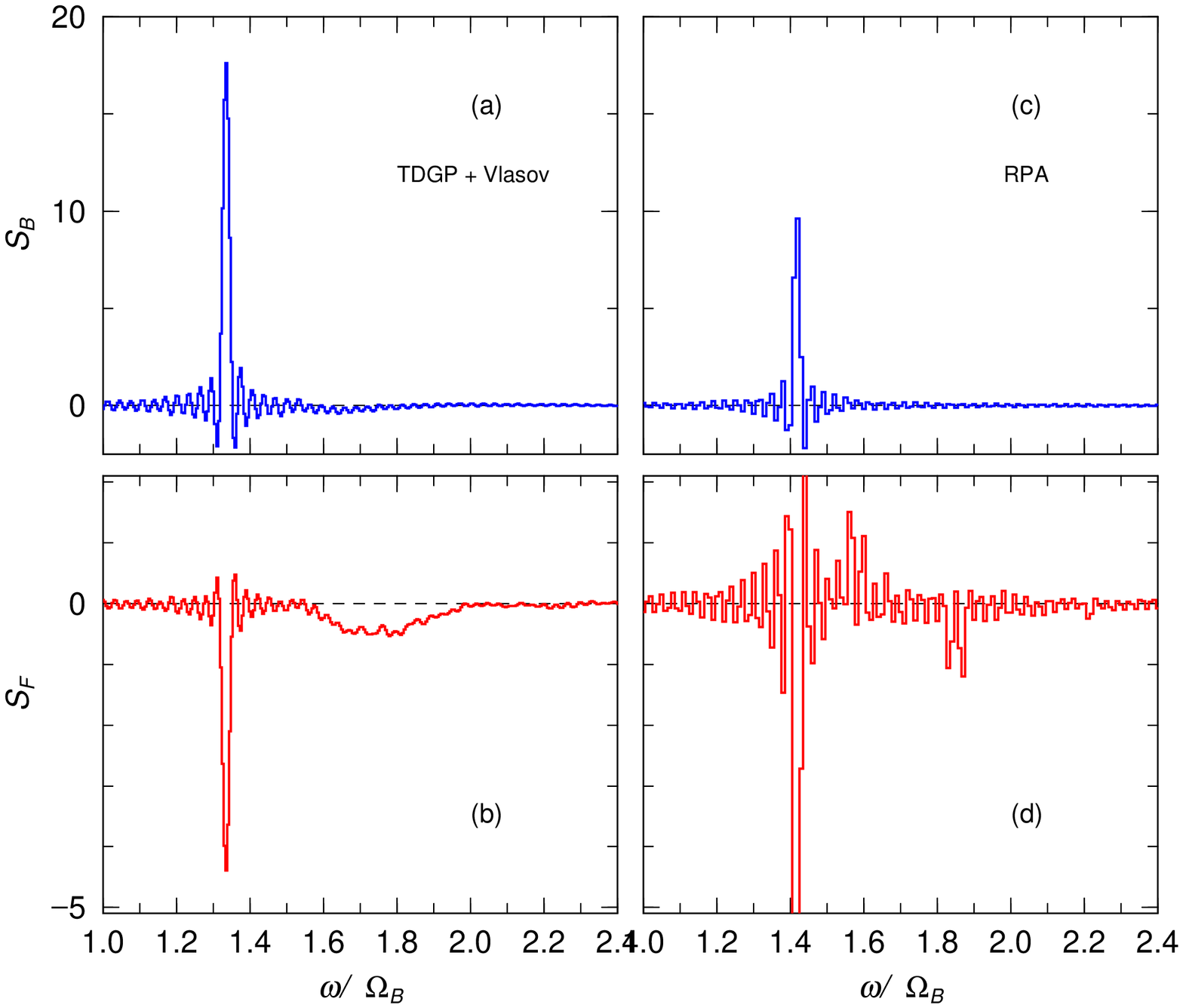}
\caption
{\small 
The strength functions of the quadrupole oscillations 
with the out-of-phase initial condition 
in {\Ybsnd} mixture (corresponding to Fig.~\ref{qYb3ou});
The panels and the lines are the same with Fig. \ref{StYb1in}.
}
\label{StYb3ou}
\end{figure}

Next, 
we show the strength functions $S_{B,F}$ in Fig.~\ref{StYb3in} 
(in-phase initial condition) 
and in Fig.~\ref{StYb3ou} (out-of-phase initial condition).
In RPA, 
the boson strengths concentrate on one sharp peak
(Fig.~\ref{StYb3in}c and Fig.~\ref{StYb3ou}c), 
and the fermion strengths have one sharp peak at the same frequency 
with the $S_B$ and have small peaks 
with broad widths
(Fig.~\ref{StYb3in}d and Fig.~\ref{StYb3ou}d);
in these results,
the $S_F$ show no strong intrinsic oscillation modes.

\begin{figure}[ht]
\vspace*{-1.5cm}
\hspace*{-1.0cm}
\includegraphics[scale=0.5,angle=270]{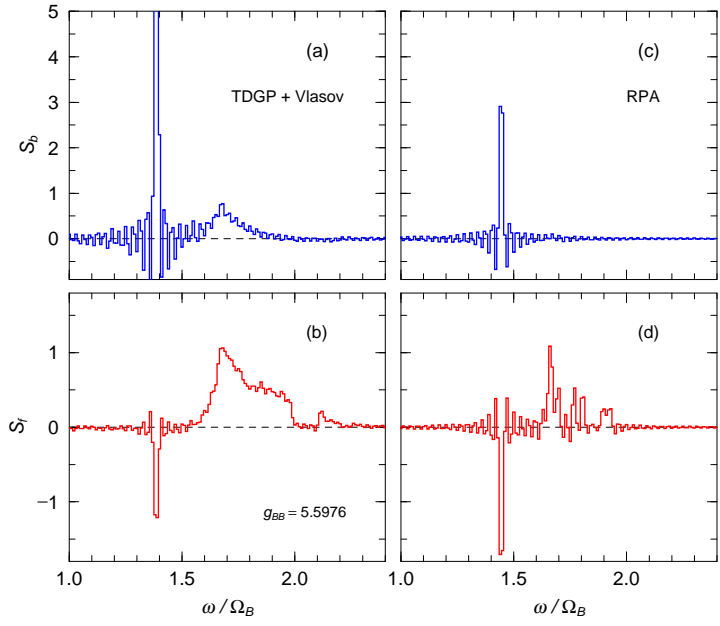}
\caption
{\small 
The strength functions of the quadrupole oscillations 
with the in-phase initial condition 
in BF mixture with $g_{BB} = 5.5976$
(other parameters are the same as {\Ybsnd} mixture);
the panels and lines are the same as in Fig.~\ref{StYb1in}.
}
\label{StYbQ3t}
\end{figure}

Here we should note that the boson-boson scattering lengths 
are very different in $^{174}$Yb and $^{170}$Yb 
as shown in TABLE I; 
thus, the boson-boson coupling constant $g_{BB}$ in \Ybthd mixture 
becomes different from that in the other mixtures.
As shown in Appendix \ref{app1}, 
in the TF approximation, 
the ground state depends only on the ratio $h_{BF}/g_{BB}$, 
and not on $g_{BB}$ directly.
In order to show this coupling-constant dependence, 
we calculate the quadrupole oscillation 
of the mixture 
with $g_{BB} = 5.598 \times 10^{-2}$ and $h_{BF} = 7.420 \times 10^{-2}$, 
which is chosen to have the same value 
of the coupling-constant ratio  
with the \Ybthd mixture $h_{BF}/g_{BB} = 1.325$.
The strength functions of the quadrupole oscillations 
are shown in Fig.~\ref{StYbQ3t}, 
which should be compared with Fig.~\ref{StYb3in} 
for the \Ybthd mixture.  
We find that hight of several peaks are different, 
but their positions are the same in these figures.
It confirms that the frequencies 
of the oscillation modes are determined 
by the ratio $h_{BF}/g_{BB}$ only 
though peak heights depend on two parameters $g_{BB}$ and $h_{BF}$.

%
%
\subsection{Fermion Intrinsic Modes and Comparison with Sum-rule Approach}
%
%

\begin{figure}[ht]
\hspace*{0cm}
{\includegraphics[scale=0.85]{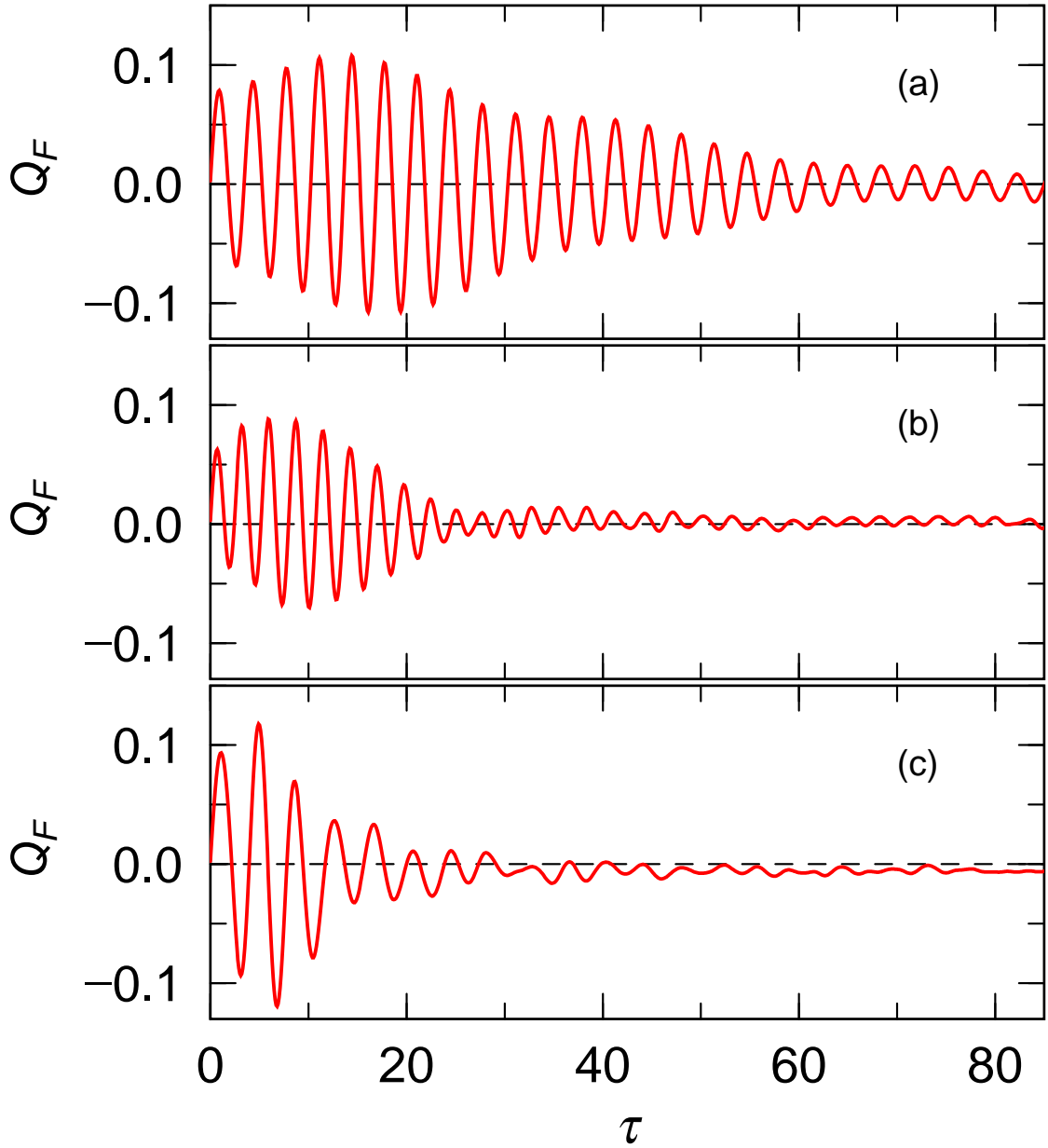}}
\caption{\small
Time evolution of the quadrupole fermion oscillations 
in $^{170}$Yb$-^{171}$Yb (a),  $^{170}$Yb$-^{171}$Yb (b) and
$^{170}$Yb$-^{171}$Yb (c), 
where the boson motions are frozen.
The initial condition is given by $\lambda_F=0.1$ for all cases.}
\label{qpFtv}
\end{figure}

In order to have some additional discussion about the fermion intrinsic modes, 
we calculate the fermion oscillation when the boson motion is frozen.
In Fig.~\ref{qpFtv}, we show the results for the mixtures of
\Ybfst (a), \Ybsnd (b) and  \Ybthd (c);
the dampings in \Ybsnd and \Ybthd are found to be much faster than that in \Ybfst.

\begin{figure}[ht]
\hspace*{0cm}
{\includegraphics[scale=0.65]{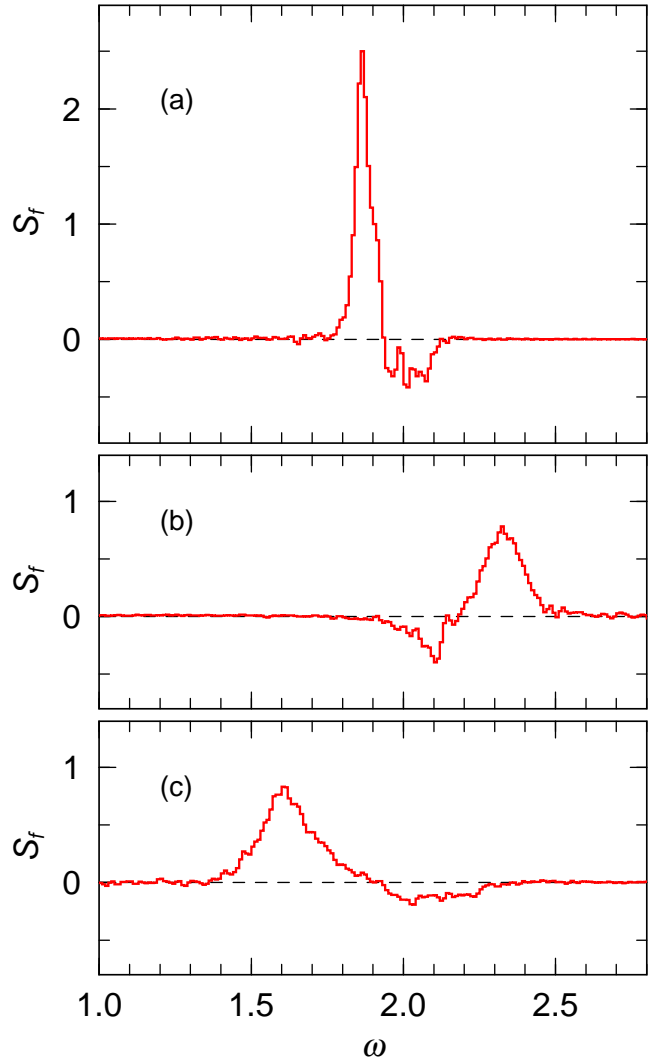}}
\caption{\small
The fermion strength functions of the fermion quadrupole oscillations 
corresponding to the cases in Fig.~\ref{qpFtv}.}
\label{StFQbf}
\end{figure}

The fermion strength functions $S_F$ of these mixtures are shown in Fig.~\ref{StFQbf}.
It is found that the $S_F$ show a large positive and a small negative peaks.  
The positions of the positive peak, 
which are different in these three mixtures, 
are at the inside-fermion oscillation frequencies (mode-3)
obtained in the previous calculations.  
The small peaks are located in 
$2.0 \le \omega \le 2.1$;  
they are supposed to be the outside-fermion oscillations (mode-2) in the previous calculations.

\begin{figure}
{\includegraphics[scale=0.85,angle=270]{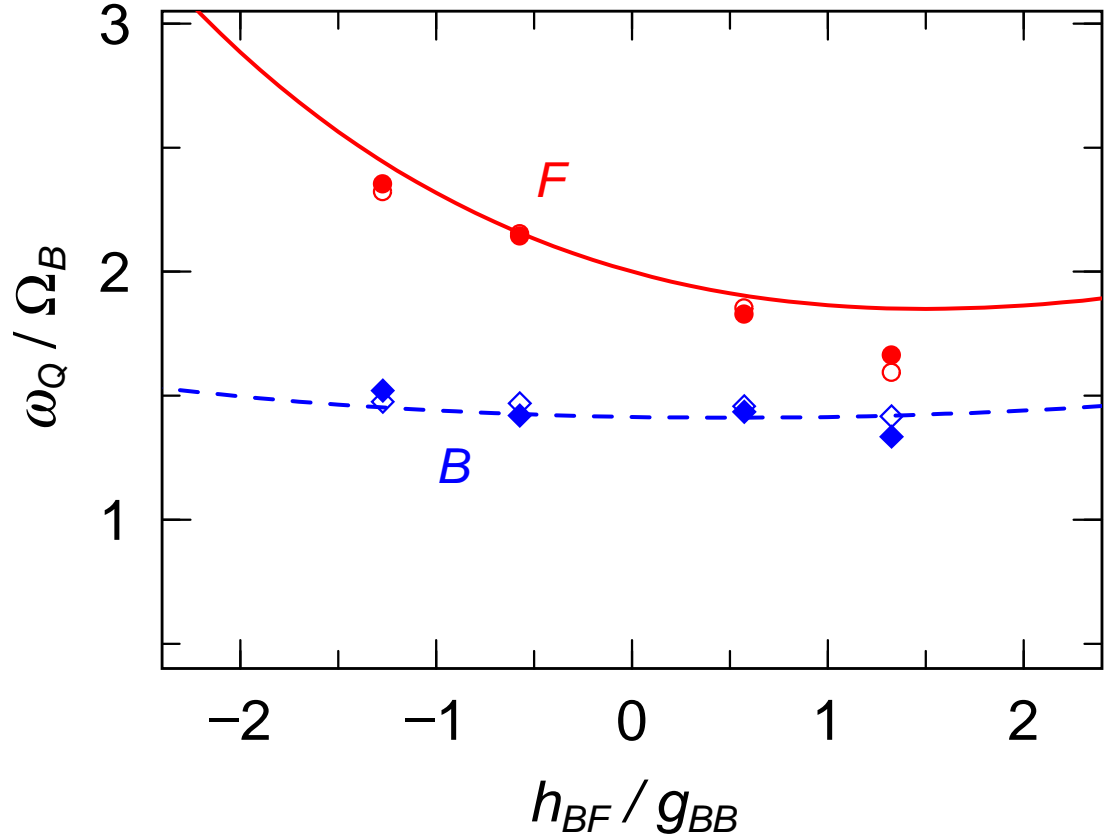}}
\caption{\small
Intrinsic frequencies of the quadrupole oscillations in boson (diamonds) 
and fermion (circles) components of the BF mixtures 
in the TDGP+Vlasov (full diamonds and circles)
and RPA (open diamonds and circles) calculations.
The solid and dashed lines are for the sum-rule results 
of the boson and fermion quadrupole modes.
}
\label{frSum}
\end{figure}

These peaks have broad widths, 
and it suggests that the oscillation modes corresponding to them have large dampings. 
Particularly, 
the outside-fermion mode has very low peak, 
so that it does not play any significant roles in the full calculations in the previous section.
As the absolute value of the  boson-fermion coupling increases, 
the effective potential for fermions has large deformation 
from the harmonic oscillator shape, 
and the unharmonic effect from this deformation makes large dampings of the fermion intrinsic modes, 
and the boson-forced oscillation mode (mode-1) have larger contribution in the full calculation. 
We show the $h_{BF}$-dependence of the boson and fermion intrinsic
frequencies in Fig.~\ref{frSum}, 
which are defined as peak positions of the strength functions.  

Now we should give some comments on the results by the sum-rule approach \cite{miyakawa}
In the sum-rule approach, 
the intrinsic frequencies of the boson and fermion 
quadrupole oscillations are obtained by
\begin{eqnarray}
     \omega_Q^b &=& \sqrt{2 - \frac{1}{N_b R_B^2}V_{QP}},
\label{sumB}\\
     \omega_Q^f &=& \sqrt{ \frac{1}{R_B^2} K_F 
                          +2 \omega_f^2 
                          -\frac{1}{m_f N_f R_F^2} V_{QP}},
\label{sumF}
\end{eqnarray}
where
\begin{eqnarray}
     K_F &=& \frac{1}{10 \pi^2 m_f^2} 
             \int d^3{r} (6 \pi \rho_F)^{\frac{5}{3}}, \\
     V_{QP} & = &- \frac{8}{5} h_{BF} \int d^3{r} r^2
                 \frac{\partial \rho_B}{\partial r} 
                 \frac{\partial \rho_F}{\partial r}.
\label{Vsum}
\end{eqnarray}
The solid and dashed lines Fig.~\ref{frSum} 
show the frequencies of the fermion and boson quadrupole oscillation 
obtained in the sum-rule approach.

The results of the sum-rule approach well reproduce 
the TDGP+Vlasov and RPA calculations, 
except the fermion intrinsic frequencies at $h_{BF}/g_{BB}=1.3256$.
It should be noted that, when $h_{BF}/g_{BB} > 1$, 
the fermion density becomes surface-peaked, and $V_{QP} < 0$, 
so that, in the sum-rule approach, 
the intrinsic frequencies take minimum values 
at $h_{BF}/g_{BB} \approx 1$ as shown in Fig.~\ref{frSum}, 
and increase $h_{BF}/g_{BB} > 1$.

As mentioned before, 
when $h_{BF}/g_{BB} > 1$,
the density- and velocity-distribution changes occur through the oscillation
in the case of the large amplitudes, 
and the potential minimum for the fermion, 
which exists at the border of the boson-distributed region, 
also changes its position.
Thus, it causes the smaller intrinsic frequency of fermion ($\omega_Q^f$) 
than the result obtained in the sum-rule approach, 
because the sum-rule approach assumes
the small amplitude oscillations, 
and the density-distribution changes through oscillation are not included.

Anyway, in the case of the repulsively-strong boson-fermion interaction,
the peak of the fermion strength function is small and broad, 
and the contribution from the fermion intrinsic mode is not so large 
except early stage in oscillation.

%
%
\section{Summary}
%
%

In this paper, 
we have investigated the collective quadrupole oscillation 
in three kinds of the BF mixtures of the Yb isotopes:
\Ybfst, \Ybsnd and \Ybthd, 
where the boson-fermion interactions are weakly repulsive, 
strongly attractive and strongly repulsive.   
In actual numerical calculations, 
we have obtained the time-evolutions of the oscillating mixtures directly
using the TDGP and Vlasov equations,  
and compare the results with the RPA calculations. 

Theoretically, 
these two approaches predict the same modes of oscillations: 
the intrinsic (mode-2,3) and boson-forced (mode-1) oscillations.
Nevertheless, the oscillation behaviors are quite different 
in these two approaches, especially in later stage of oscillation.

When the boson-fermion interaction is weak, 
the two approaches give almost the same results in early stage, 
but some difference appears in latter stage.
In the mixture with very strong BF interaction, 
the difference appears also in the earlier stage;
in \Ybsnd and \Ybthd mixtures, 
the two approaches are consistent
only in the first one or two periods of the oscillations. 

In the case of the strongly-attractive BF interaction (\Ybsnd), 
the fermion overflow from the boson-distributed region into the outside
causes the fermion gas expansion. 
On the other hand, When the BF interaction is strongly repulsive (\Ybthd), 
the fermion oscillation loses the strength of its intrinsic mode soon, 
and the fermi gas oscillates with the same period of the boson gas
in later period.

The RPA is available only in the case of the small-amplitude oscillations, 
because it cannot trace the density-distribution changes in the course of time evolution.
In actual experiments, 
the amplitude is not so small, 
and the methods in solving the time-dependence process 
should be proper in comparison with experiments.  

In this paper, 
we assume the spherical trapping potential with $\kappa_L=1$, 
but the actual experiments will be done with the largely-deformed potential 
with $\kappa_L=1/6$, for example, in Kyoto group.
In such cases, 
the breathing oscillations are not decoupled into the monopole and quadrupole modes \cite{scalTF2}
but into the longitudinal and transverse oscillation modes.
The results of the collective oscillations of the BF mixtures 
in the deformed trapping potential will be discussed in another paper \cite{MYnext}.

Furthermore, we do not take into account two-body collisions and thermal
boson effects \cite{JackZar}.
In the system $N_b \gg N_f$ at $T=0$, 
the number of thermal bosons are very small, 
so that the two-body collisions are not expected to play any significant roles 
in the oscillation processes. 
In the actual experiments, which are performed at very low but $T \neq 0$ temperatures, 
thermal bosons should give some contributions; 
the introduction of such effects through two-body collision terms into our approach
should be done in the future \cite{BUU1,TOMO1}.

\newpage

\appendix

\section{Ground State of Boson-Fermion Mixtures in Thomas-Fermi
 Approximation}
\label{app1}

In this appendix, 
we briefly explain the ground state of the BF mixture 
in the Thomas-Fermi (TF) approximation.

In this approximation, 
the total energy of the BF mixture is given by
\begin{eqnarray}
     E_T = \int d^3{r} \Bigg\{
               & & \frac{1}{2} (\vbr_T^2 +\kappa_L^2 r_3^2) \rho_B(\vbr)
               +\frac{g_{BB}}{2} \rho_B^2(\vbr)  
\nonumber\\
               &+& \frac{1}{20 \pi^2 m_f} [6 \pi \rho_F (\vbr)]^{5/3}
                + \frac{1}{2} m_f \omega_f^2 
                   (\vbr_T^2 + \kappa_L^2 r_3^2) \rho_F(\vbr) 
\nonumber\\
               &+& h_{BF}  \rho_B  (\vbr)  \rho_F(\vbr) \Bigg\},
\end{eqnarray}

The scaled dimensionless variables and parameters are defined by
\begin{eqnarray}
     h = \frac{1}{m_f \omega_f^2} \frac{h_{BF}}{g_{BB}},  &\qquad&
    \vx = \frac{m_f^4 \omega_f^5 g_{BB}}{3 \pi^2} 
          ( r_1, r_2, \kappa_L r_3 ),
\nonumber\\
     n_B = \frac{2 m_f^8  \omega_f^{10} g_{BB}^3}{9 \pi^4} \rho_B, &\qquad&
     n_F = \frac{2 m_f^9  \omega_f^{12} g_{BB}^3}{9 \pi^4} \rho_F,
\nonumber\\
     e_B = \frac{2 m_f^8  \omega_f^{10} g_{BB}^2}{9 \pi^4} \mu_B, &\qquad&
     e_F = \frac{2 m_f^7  \omega_f^{8} g_{BB}^2}{9 \pi^4} \mu_F.
\end{eqnarray}
The effective boson and fermion numbers are defined by
\begin{eqnarray}
     \Nb &=& \int d^3 x n_B 
          =  \frac{2 \kappa_L m_f^{20} \omega_f^{25} g_{BB}^6}{3^5 \pi^{10}} N_B,
\\ 
     \Nf &=& \int d^3 x n_F 
          =  \frac{2 \kappa_L m_f^{21} \omega_f^{27} g_{BB}^6}{3^5 \pi^{10}} N_F.
\end{eqnarray}
Finally, the scaled total energy becomes
\begin{eqnarray}
     \Et &=& \frac{2 \kappa_L m_f^{28} \omega_f^{35} g_{BB}^8}{3^7 \pi^{14}} E_T
\nonumber \\
         &=& \int d^3{x} \left\{ x^2 n_B  +\frac{1}{2} n_B^2 
                                + \frac{3}{5} n_F^{\frac{5}{3}} 
                                + x^2 n_F + h n_B n_F \right\},
\label{etotTFsg}
\end{eqnarray}
where $x^2 = |\vx|^2$. 

Let's introduce the particle-number constraints into the total energy 
as ${\tilde E}' = \Et - e_B \Nb - e_F \Nf$, 
where the Lagrange multipliers $e_B$ and $e_F$ are the scaled boson and fermion chemical potentials.
The variations of ${\tilde E}'$,
$\delta {\tilde E}^{\prime}/\delta n_B = 0$ and 
$\delta {\tilde E}^{\prime}/\delta n_F = 0$, 
gives the TF equations for the ground-state densities $n_{B,F}$¡§
\begin{eqnarray}
     n_B + h n_F &=& e_B - x^2 ,
\label{TFbs}\\
     n_F^{\frac{2}{3}} + h n_B &=& e_F -x^2,
\label{TFfr}
\end{eqnarray}
The second order variations give 
the stability condition of the TF ground states:
\begin{equation}
     \frac{\delta^2 \Et}{\delta n_B^2} 
     \frac{\delta^2 \Et}{\delta n_F^2}
          -\left(\frac{\delta^2 \Et}{\delta n_B \delta n_F}\right)^2 > 0,
\end{equation}
which leads to
\begin{equation}
     n_F^{1/3}  < \frac{2}{3 h^2}.  
\label{stCon}
\end{equation}

It should be noted that the scaled equations include three parameters, 
$(e_B, e_F, h)$
though four parameters $(\mu_B, \mu_F, g_{BB}, h_{BF})$ 
exist in the original unscaled system.

Using $s_F = [n_F(x)]^{1/3}$ and eliminating $n_B$ in Eqs.~(\ref{TFbs}, \ref{TFfr}),
we obtain 
\begin{equation}
     f(s_F) = s_F^2 - h^2 s_F^3 - (e_F - h e_B) + (1-h)x^2 = 0.
\label{FtfEq}
\end{equation}
From the derivative function of (\ref{FtfEq}):
\begin{equation}
     \frac{df}{d s_F} = 2 s_f - 3 h^2 s_F^2 
                      = 3h^2 s_F^2 \left( \frac{2}{3h^2} - s_F \right),
\end{equation}
we find that $d f/d s_F > 0$ for $s_F < 2/3h^2$;
it is exactly equivalent to the stability condition in (\ref{stCon}).
It means the existence of the solution with the positive value in (\ref{FtfEq}) 
when $f(0) < 0 < f(2/3h^2)$.

Now we consider the condition that 
the fermion density has a maximum peak at the surface.
Differentiate Eq.~(\ref{FtfEq}) with respect to $x$,
we obtain
\begin{equation}
     3h^2 s_F^2 \left(\frac{2}{3h^2} - s_F\right) 
     \frac{\partial s_F}{\partial x} = -2 (1-h)x,
\label{dTFfr}
\end{equation}
Using the stability condition in (\ref{stCon}), 
Eq.~(\ref{dTFfr}) gives  $\partial s_F/\partial x  < 0$ when $h<1$ 
(case 1),
and $\partial s_F/\partial x  > 0$ when $h>1$ 
(case 2); 
in case 1, the TF fermion density $n_F$ should have the maximum 
at $x=0$ (the center-peaked profile), 
and, in case 2, $x=0$ should be the minimum of the $n_F$, 
which should have a maximum outside the boson occupation region.
(the surface-peaked profile).

An extreme case of the surface-peaked fermion densities ($h >1$) 
is the shell-structure profile, 
where fermions are pushed outside and no fermions exist in the central region. 
The shell-structure profile should appear when the TF solution  
satisfies $n_F(0) < 0$; using Eq.~(\ref{FtfEq}), 
we find that it occurs when $h e_B > e_F > e_B$. 
It should be noted that, when $e_B > e_F$, 
there are no solutions in Eq.~(\ref{FtfEq}).
  
When  $h < 1$ (the center-peaked fermion profile),
the fermion density at the center is required to be positive, 
$n_F(0) > 0$; otherwise there are no solutions in Eq.~(\ref{FtfEq}).
The condition, $n_F(0) > 0$, gives the restriction of the parameters as
follows:
\begin{equation}
he_B < e_F < h e_B + \frac{4}{27 h^4} .
\end{equation}

When $0<h<1$,
there exists the case that the boson density also shows the surface-peaked profile.
In order to obtain the second derivative of $n_B$ at $x=0$, 
we two-time differentiate Eqs.~(\ref{TFbs}, \ref{TFfr}) with respect to $x$:
\begin{eqnarray}
     \left. \frac{\partial^2 n_B}{\partial x^2} \right|_{x=0} 
      +h \frac{\partial^2 n_F}{\partial x^2}|_{x=0} &=& -2,
\label{TFbssd}
\\
     \frac{2}{3} [n_F (0)]^{-\frac{1}{3}} 
     \frac{\partial^2 n_F}{\partial x^2}|_{x=0}
          +h \left. \frac{\partial^2 n_B}{\partial x^2} \right|_{x=0} &=& -2.
\label{TFfrsd}
\end{eqnarray}
Solving the above equations for $\partial^2 n_B/\partial x^2$, 
we obtain
\begin{equation}
     \frac{\partial^2 n_B}{\partial x^2}|_{x=0} 
          = \left\{ -2 + 3 h [n_F(0)]^{\frac{1}{3}} \right\} 
            \left\{ 1 -\frac{3}{2} h^2 [n_F(0)]^{\frac{1}{3}} \right\}^{-1}.
\end{equation}
It shows that, when $ 2/3h < s_F(0) =[n_F(0)]^{1/3} < 2/3h^2$, 
$\partial^2 n_B/\partial x^2 > 0$ at $x=0$;
the boson density have the surface-peaked profile
(It should be noted that $h$ satisfies $2/3h < 2/ 3 h^2$
because we consider the case of $0 < h < 1$).
The condition that the solution of Eq.~(\ref{FtfEq}) satisfies 
$2/3h < s_F(0) <2/3h^2$ is obtained by $f(2/3h) < 0 < f(2/3h^2)$ at $x=0$;
using Eq.~(\ref{FtfEq}), 
we obtain the condition that the boson density has the surface-peaked profile:
\begin{equation}
    h e_B +\frac{4}{9 h^2} -\frac{8}{27 h} < e_F < h e_B  + \frac{4}{27 h^4} . 
\end{equation}
Also, $n_B(0) < 0$ gives the boson shell-structure condition: $e_B < h e_F^{3/2}$.

\newpage
\begin{table}[ht]
\begin{tabular}{|c|c|cc|cc|c|}
\hline
~~~~~& System & ~~$a_{BB}$ (nm)~~ & ~~$(100g_{BB})$~~ & ~~$a_{BF} $ (nm)~~ &  
~~$(100h_{BF})$~~ & ~~ $h_{BF}/g_{BB}$ ~~
\\
\hline 
(1)~& ~~$^{170}$Yb $-$ $^{171}$Yb~~ & ~~ 3.4353~~ &( 5.5976 )~~ & ~1.9680
 & (~3.2067)  & ~~~0.573~ 
\\
\hline
(2)~&~~$^{170}$Yb $-$ $^{173}$Yb~~ & ~~ 3.4353~~ &( 5.5976 )~~ & -4.3730 
& (-7.1255) & ~-1.273~ 
\\
\hline
(3)~&~~$^{174}$Yb $-$ $^{173}$Yb~~ & ~~ 5.4630~~ &( 8.9016 )~~ & ~7.2410
 & (11.7994) & ~~1.325~  
 \\
\hline
\end{tabular}
\caption{\small
Scattering lengths (coupling constant) of the BF mixtures 
of the Yb isotopes.
$a_{BB,BF}$ ($g_{BB}$,$h_{BF}$) are the boson-boson and boson-fermion 
scattering lengths (coupling constants). }
\label{coupTB}
\end{table}

\vspace*{1cm}

\clearpage

\newpage

\clearpage

\newpage

\clearpage

\newpage

\newpage

\clearpage

\newpage

\newpage

\newpage

\newpage

\newpage

\newpage

\newpage

\newpage

\clearpage

\newpage

\end{document}